\documentclass[aps,pra,12pt]{revtex4}
\usepackage{mathrsfs}
\usepackage{mathbbold}
\usepackage{amssymb}
\usepackage{amsfonts}
\usepackage{amsmath}
\usepackage{amscd}
\usepackage{graphicx}
\usepackage{cases}

\begin{document}

\title[Short Title]{Heralded atomic nonadiabatic holonomic quantum computation with Rydberg blockade}

\author{Yi-Hao Kang$^{1,2}$}
\author{Zhi-Cheng Shi$^{1,2}$}
\author{Jie Song$^{3}$}
\author{Yan Xia$^{1,2,}$\footnote{E-mail: xia-208@163.com}}

\affiliation{$^{1}$Fujian Key Laboratory of Quantum Information and Quantum Optics (Fuzhou University), Fuzhou 350116, China\\
             $^{2}$Department of Physics, Fuzhou University, Fuzhou 350116, China\\
             $^{3}$Department of Physics, Harbin Institute of Technology, Harbin 150001, China}

\begin{abstract}

We propose a protocol to realize atomic nonadiabatic holonomic
quantum computation (NHQC) with two computational atoms and an
auxiliary atom. Dynamics of the system is analyzed in the regime of
Rydberg blockade, and robust laser pulses are designed via reverse
engineering, so that quantum gates can be easily realized with high
fidelities. In addition, we also study the evolution suffering from
dissipation with a master equation. The result indicates that decays
of atoms can be heralded by measuring the state of the auxiliary
atom, and nearly perfect unitary evolution can be obtained if the
auxiliary atom remains in its Rydberg state. Therefore, the protocol
may be helpful to realize NHQC in dissipative environment.

\end{abstract}

\maketitle

\section{Introduction}

Quantum computation, by means of coherence superpositions of quantum
states, has shown many advantages in solving some problems such as
factoring and searching
\cite{Shormetting,GroverPRL79,LongPRA64,ParedesPRL95}. To accurately
execute quantum computation, high-fidelity quantum gates are
essential. However, in the implementations of quantum gates with
realistic physical systems, imperfections like systematic errors,
classical parameter fluctuation and dissipation, which all limit
gate fidelities. Therefore, how to eliminate or reduce influence of
these imperfections is a critical problem in the realization of
practical quantum computation.

In recent years, concept of nonadiabatic geometric quantum
computation (NGQC)
\cite{HerterichPRA94,ZSLPRL91,WZSPRA76,FriedenauerPRA67,CTPRApp10,XGFPRA92}
based on Abelian-geometric-phase and nonadiabatic holonomic quantum
computation (NHQC)
\cite{SjoqvistNJP14,XGFPRL109,XGFPRA952,ZPZPRA94,XZYPRA92,KYHADP531}
based on non-Abelian-geometric-phase have been proposed in order to
enhance robustness against experimental imperfections. Thereinto,
NHQC have shown advantages in various ways. Firstly, as geometric
phases are determined by global properties of evolution paths, they
are insensitive to classical parameter fluctuation over cyclic
evolution
\cite{BerryPRSA392,AharonovPRL58,SjoqvistPhys1,LQXPRA101,ZJSR5,ZSLPRA72,ZZPRApp12,ZPZPRA95}.
Thus, NHQC can be well performed in the presence of classical
parameter fluctuation \cite{GuridiPRL94,ZJPRA97}. Secondly, compared
with traditional adiabatic holonomic quantum computation (AHQC)
\cite{ZanardiPLA264,WLAPRL95,DLMSci292,PachosPRA61}, NHQC releases
variations of parameters from limitation of the adiabatic condition,
consequently reducing exposure of physical systems to dissipation.
Thirdly, recent researches \cite{LBJPRL123,LSAQT3,KYHPRA101,DYXAQT2}
have shown that NHQC can be implemented in a flexible way,
compatible with many control and optimal methods, e.g., reverse
engineering
\cite{VitanovPRA85,GaraotPRA89,LYCPRA97,KYHPRA97,CXPRA83,LYCNJP20,KYHPRA100,RousseauxPRA87,RDPRA101}
and systematic-error optimal method
\cite{RuschhauptNJP14,YXTPRA97,DaemsPRL111,DammePRA96}. Therefore,
robustness of NHQC against systematic errors can be greatly improved
by combining with proper techniques. To date, take advantage of
NHQC, many robust protocols for quantum computation
\cite{MousolouPRA89,ZJPRA89,XGFPRA95,XZYPRApp7,LZTPRA89,SXKNJP18,XZYPRA94,HBHADP530}
have been put forward. Moreover, stability of NHQC has also been
demonstrated in a number of experiments
\cite{AbdumalikovNat496,ZCNat514,CamejoNC5,ZBBPRL119,XYPRL121,NagataNC9}.

In the implementation of NHQC, time required is shortened compared
with AHQC, and quantum information can be encoded in
decoherence-free subspace. However, lossy intermediate states, such
as excited states of atoms and states of non-vacuum cavity, may
still be required as auxiliaries
\cite{ZPZPRA98,ZPZPRA99,KYHPRA972,HZPPRA97,LBJPRA95}. During
operations, dissipation acting on these intermediate states, such as
energy relaxation and cavity photon leakage, also spoils the unitary
evolution. As a result, a system initially in a pure state would
finally be in a mixed state after operations, and fidelities of NHQC
decrease simultaneously. To further improve fidelities of NHQC for
constructions of functional quantum computers, more efforts should
be made to diminish the impact of dissipation. Fortunately,
researches in past few years have shown some interesting ways to
overcome the influence of dissipation in quantum evolutions.
Especially, one approach suggests to design physical systems in
order that decays of qubits can be reported by measuring auxiliary
qubits, which have been successfully used in protocols
\cite{ShwaPRA88,GarttnerPRA92,JohnsonPRL109,CYAPRL104} for
generating pure entangled states, namely the heralded entanglement
generations. More interestingly, the heralded protocols can even be
extended to the implementations of quantum gates
\cite{BorregaardPRL114,QWPRA96}, where perfect unitary evolutions
can be maintained in dissipative environment if correct results are
reported in measurements of auxiliary qubits. The inspiring ideas of
the heralded protocols make us to think whether they can be
incorporated in NHQC, so that quantum computation can be implemented
with comprehensive resistance to systematic errors, classical
parameter fluctuation and dissipation.

In this paper, we present a protocol to realize heralded NHQC with a
physical system containing two computational atoms and an auxiliary
atom. With the help of Rydberg blockade, we restrict the evolution
in a twelve-dimensional subspace. Afterwards, the evolution is
further studied by invariant-based reverse engineering, where paths
for NHQC are naturally constructed by eigenvectors of an invariant.
Moreover, the parameters are meticulously selected by nullifying
systematic error sensitivity. Numerical results demonstrate that the
implementations of quantum gates are insensitive to systematic
errors of laser pulses. Furthermore, dissipation is also taken into
account. By analyzing the evolution governed by a master equation,
we show that decays of atoms can be reported by measuring the state
of the auxiliary atom after operations, and the unitary operation
can be maintained if the auxiliary atom is still in its Rydberg
state. Compared with the previous Rydberg-atom-based NHQC protocol
\cite{KYHPRA972}, in which the system may become in a mixed state in
the presence of dissipation, the system in the current protocol can
remain in a pure state with correct measurement result. Last but not
least, the protocol also benefits a lot from merits of Rydberg atoms
and Rydberg blockade. For example, long-live time of Rydberg states
\cite{MullerPRA89,SaffmanJPB49,SSLPRA93} is helpful to produce high
successful probability of the protocol. Moreover, Rydberg blockade
is also great help to reduce mechanical effect and ionization
\cite{BurkhardtPRA34,VrinceanuPRA72,RobicheauxPRA56,FreitagPRB95}
which play significant roles when multiple atoms are excited to
their Rydberg states. Overall, the protocol shares advantages of
geometric phase, reverse engineering, systematic-error-sensitivity
nullified optimal control, heralded implementation and stability of
Rydberg states. Therefore, the protocol may provide useful
perspectives in the realization of high-fidelity quantum
computation.

The article is organized as follows. In Sec. II, we review the
theories for realizing NHQC with invariant-based reverse
engineering. In Sec. III, we give the effective Hamiltonian of an
atomic system and study the evolution under the influence of
dissipation. In Sec. IV, we amply discuss implementations of
arbitrary single-qubit gates with invariant-based reverse
engineering, and select robust parameters by nullifying the
systematic error sensitivity. In Sec. V, we describe the
invariant-based implementations of two-qubit entangling gates. In
Sec. VI, we perform numerical simulations to check the performance
of the implementations of single- and two-qubit gates. Finally,
conclusions are given in Sec. VII.

\section{Theoretical preparation}

\subsection{Lewis-Riesenfeld invariant theory}

Let us first briefly review the Lewis-Riesenfeld invariant theory
\cite{LewisJMP10}. We assume that a Hermitian operator $I(t)$ obey
the equation as $(\hbar=1)$
\begin{equation}\label{lr1}
i\frac{\partial}{\partial t}I(t)-[H(t),I(t)]=0,
\end{equation}
with $H(t)$ being the Hamiltonian of the considered physical system.
If $|\phi_l(t)\rangle$ is a non-degenerate eigenvector of $I(t)$,
one can derive a solution of the time-dependent Schr\"{o}dinger
equation $i|\dot{\psi}(t)\rangle=H(t)|\psi(t)\rangle$ as
$|\psi_l(t)\rangle=\exp[i\alpha_l(t)]|\phi_l(t)\rangle$, where
$\alpha_l(t)$ is the Lewis-Riesenfeld phase for $|\phi_l(t)\rangle$
defined as
\begin{equation}\label{lr2}
\alpha_l(t)=\int_{0}^{t}\langle\phi_l(t')|i\frac{\partial}{\partial
t'}-H(t')|\phi_l(t')\rangle dt'.
\end{equation}
Therefore, the dynamic invariant $I(t)$ can help us to analyze the
evolution of the system. In practice, a very useful way to construct
dynamic invariant is to use Lie algebra
\cite{GungorduPRA86,TorronteguiPRA89}. After constructing a dynamic
invariant $I(t)$, by making proper ansatz for parameters of $I(t)$,
one can reversely derived Hamiltonian $H(t)$ via Eq.~(\ref{lr1})
\cite{CXPRA86}. We will make some introductions about
Lie-algebra-based construction of invariants and reverse engineering
of Hamiltonian in Sec. IIB.

\subsection{Construction of invariants and reverse engineering of Hamiltonian by using Lie algebra}

We consider a system with Hamiltonian $H(t)$, which can be expressed
by Hermitian generators $\{G_m|m=1,2,...,\tilde{m}\}$ (satisfying
orthogonal condition with the Hilbert-Schmidt inner product
$(G_m,G_m')=\mathrm{Tr}(G_mG_{m'}^\dag)=0$, $m\neq m'$) of a Lie
algebra $\mathcal{G}$ (dynamical algebra \cite{KaushalJMP22}) as
$H(t)=\sum_{m=1}^{\tilde{m}}h_m(t)G_m$ with $\{h_m(t)\}$ being real
parameters. In addition, we assume an invariant $I(t)$ in form of
$I(t)=\sum_{m=1}^{\tilde{m}}\xi_m(t)G_m$ with $\{\xi_m(t)\}$ being
real parameters. Thus, both $H(t)$ and $I(t)$ can be deemed as
vectors in $\mathcal{G}$ with basis $\{G_m\}$. Defining a linear
transformation $\mathscr{F}_m:\mathcal{G}\mapsto\mathcal{G}$ related
to $G_m$ and acting on arbitrary vector $X\in\mathcal{G}$ as
$\mathscr{F}_mX=-i[G_m,X]$, we can construct linear transformations
$\mathcal{G}\mapsto\mathcal{G}$ related to the Hamiltonian $H(t)$
and the invariant $I(t)$ as
$\mathscr{H}(t)=\sum_{m=1}^{\tilde{m}}h_m(t)\mathscr{F}_m$ and
$\mathscr{I}(t)=\sum_{m=1}^{\tilde{m}}\xi_m(t)\mathscr{F}_m$,
respectively. Thus, Eq.~(\ref{lr1}) can be rewritten by
$\dot{I}(t)=\mathscr{H}(t)I(t)$, which is a set of first order
linear differential equations. According to the existence theorem of
solutions for first order linear differential equations
\cite{AhmadBook}, one can find a solution for $I(t)$ theoretically.
Therefore, the assumption $I(t)=\sum_{m=1}^{\tilde{m}}\xi_m(t)G_m$
is proper.

Although one can always find an invariant in the dynamical algebra
with given Hamiltonian parameters $\{h_m(t)\}$ in theory, the
process of obtaining analytical solution of $I(t)$ with
$\dot{I}(t)=\mathscr{H}(t)I(t)$ is usually very complex. Thus,
instead of solving the invariant $I(t)$ with known $\{h_m(t)\}$,
reverse engineering suggests to reversely derive Hamiltonian of the
system by using parameters $\{\xi_m(t)\}$. For this sake, we rewrite
Eq.~(1) as $\dot{I}(t)=\mathscr{I}(t)H(t)$. Since
$\mathscr{I}(t)I(t)=0$, $\mathscr{I}(t)$ is a singular matrix,
consequently $\mathscr{I}^{-1}(t)$ does not exist. Thus, $H(t)$ can
not be solved by $H(t)=\mathscr{I}^{-1}(t)I(t)$. In practice, one
can apply Gauss elimination or pseudo-inverse matrices
\cite{TorronteguiPRA89} to solve $H(t)$. Here, we take Gauss
elimination as example to illustrate the process of reverse
engineering. Noticing that some of controls in $\{G_m\}$ may be
unavailable in some specific systems, we firstly partition
$\mathscr{I}(t)$ and $H(t)$ as
\begin{eqnarray}\label{plr6}
\mathscr{I}(t)=\left[%
\begin{array}{cc}
  \mathscr{I}^{(1)}_{\tilde{m}\times \tilde{m}_1}(t) &\ \mathscr{I}^{(2)}_{\tilde{m}\times \tilde{m}_2}(t)
\end{array}%
\right],\ H(t)=\left[%
\begin{array}{c}
  H_{\tilde{m}_1}(t) \\
  O_{\tilde{m}_2}(t) \\
\end{array}%
\right],
\end{eqnarray}
with $\tilde{m}_1$ and $\tilde{m}_2$ being numbers of available and
unavailable controls, respectively. In this case, we obtain equation
as $\mathscr{I}^{(1)}_{\tilde{m}\times
\tilde{m}_1}(t)H_{\tilde{m}_1}(t)=\dot{I}$. We assume that the rank
of $\mathscr{I}^{(1)}_{\tilde{m}\times \tilde{m}_1}(t)$ is
$\tilde{m}_3$ ($\tilde{m}_3\leq\tilde{m}_1$ is always satisfied due
to the property of rank). Thus, one can transform
$\mathscr{I}^{(1)}_{\tilde{m}\times \tilde{m}_1}(t)$ to
\begin{eqnarray}\label{plr7}
\mathscr{\tilde{I}}^{(1)}_{\tilde{m}\times \tilde{m}_1}(t)=\mathscr{R}(t)\mathscr{I}^{(1)}_{\tilde{m}\times \tilde{m}_1}(t)=\left[%
\begin{array}{cc}
  \mathscr{\tilde{I}}^{(1)}_{\tilde{m}_3\times \tilde{m}_1}(t) \\
  O_{\tilde{m}_4\times \tilde{m}_1}(t)
\end{array}%
\right],
\end{eqnarray}
by a set of elementary row transformations as $\mathscr{R}(t)$ with
$\tilde{m}_4=\tilde{m}-\tilde{m}_3$. Then, we can partition
$\dot{\tilde{I}}(t)=\mathscr{R}(t)\dot{I}(t)$ as
$\dot{\tilde{I}}(t)=[\dot{\tilde{I}}_{\tilde{m}_3}(t)\
\dot{\tilde{I}}_{\tilde{m}_4}(t)]^T$, and derive
$\mathscr{\tilde{I}}^{(1)}_{\tilde{m}_3\times
\tilde{m}_1}(t)H_{\tilde{m}_1}(t)=\dot{\tilde{I}}_{\tilde{m}_3}(t)$
and $\dot{\tilde{I}}_{\tilde{m}_4}(t)=0$. Thereinto,
$\dot{\tilde{I}}_{\tilde{m}_4}(t)=0$ is a set of constraint
equations for parameters $\{\xi_m(t)\}$. Thus, the number of
independent parameters in $\{\xi_m(t)\}$ is $\tilde{m}_3$. In
addition, because of $\tilde{m}_3\leq\tilde{m}_1$, solutions of
equation $\mathscr{\tilde{I}}^{(1)}_{\tilde{m}_3\times
\tilde{m}_1}(t)H_{\tilde{m}_1}(t)=\dot{\tilde{I}}_{\tilde{m}_3}(t)$
always exist. When $\tilde{m}_1=\tilde{m}_3$, $H_{\tilde{m}_1}(t)$
have a unique solution
$[\mathscr{\tilde{I}}^{(1)}_{\tilde{m}_3\times
\tilde{m}_1}(t)]^{-1}\dot{\tilde{I}}_{\tilde{m}_3}(t)$. When
$\tilde{m}_1>\tilde{m}_3$, one can obtain multiple solutions of
$H_{\tilde{m}_1}(t)$, and the number of linearly independent
solutions is $(\tilde{m}_1-\tilde{m}_3)$.

\subsection{Nonadiabatic holonomic quantum computation with eigenvectors of dynamic invariant}

To realize nonadiabatic holonomic quantum computation (NHQC) in a
computational subspace $\mathcal{S}$, an alternative way is to
select a set of time-dependent vectors
$\{|\tilde{\psi}_l(t)\rangle\}$ spanning $\mathcal{S}$ and meeting
the cyclic evolution condition
$|\tilde{\psi}_l(0)\rangle=|\tilde{\psi}_l(T)\rangle$ ($T$ is the
total operation time). According to Ref.~\cite{LBJPRL123}, if the
operator
$\tilde{\Xi}_l(t)=|\tilde{\psi}_l(t)\rangle\langle\tilde{\psi}_l(t)|$
obeys the von Neumann equation
\begin{equation}\label{lr3}
\frac{d}{dt}\tilde{\Xi}_l(t)=-i[H(t),\tilde{\Xi}_l(t)],
\end{equation}
the evolution in subspace $\mathcal{S}$ can be described as
\begin{equation}\label{lr4}
U(T,0)=\sum\limits_le^{i[\tilde{\vartheta}_l(T)+\tilde{\Theta}_l(T)]}\tilde{\Xi}(0),
\end{equation}
with
\begin{equation}\label{lr5}
\tilde{\vartheta}_l(t)=-\int_0^{t}\langle\tilde{\psi}_l(t')|H(t')|\tilde{\psi}_l(t')\rangle
dt',
\end{equation}
and
\begin{equation}\label{lr6}
\tilde{\Theta}_l(t)=\int_0^{t}\langle\tilde{\psi}_l(t')|i\frac{\partial}{\partial
t'}|\tilde{\psi}_{l}(t')\rangle dt'.
\end{equation}
being the dynamic phase and the geometric phase acquired by
$|\tilde{\psi}_l(t)\rangle$ during the time interval $[0,T]$.
Therefore, the evolution become purely geometric when
$\vartheta_l(T)=0$ for all vectors in
$\{|\tilde{\psi}_l(t)\rangle\}$. In practice, eigenvectors of a
dynamic invariant is an alternative candidate to construct vectors
$\{|\tilde{\psi}_l(t)\rangle\}$. According to Ref.~\cite{KYHPRA101},
for a non-degenerate eigenvector $|\phi_l(t)\rangle$ of a dynamic
invariant, the von Neumann equation
\begin{equation}\label{lr7}
\frac{d}{dt}\Xi_l(t)=-i[H(t),\Xi_l(t)],
\end{equation}
with $\Xi_l(t)=|\phi_l(t)\rangle\langle\phi_l(t)|$ is naturally
satisfied. Therefore, if a set of non-degenerate eigenvectors
$\{|\phi_l(t)\rangle\}$ of dynamic invariant span the computational
subspace $\mathcal{S}$, the condition to realize the NHQC is to
eliminate the dynamic part of the Lewis-Riesenfeld phase acquired in
$[0,T]$ as
\begin{equation}\label{lr8}
\vartheta_l(T)=-\int_0^{T}\langle\phi_l(t)|H(t)|\phi_l(t)\rangle
dt=0.
\end{equation}
Accordingly, the remaining part of Lewis-Riesenfeld phase
\begin{equation}\label{lr9}
\Theta_l(T)=\int_0^{t}\langle\phi_l(t)|i\frac{\partial}{\partial
t}|\phi_{l}(t)\rangle dt,
\end{equation}
is pure geometric.

\section{Dynamics of the atomic system in the Rydberg blockade regime}

\subsection{Physical model and Hamiltonian}

\begin{figure}
\scalebox{0.6}{\includegraphics[scale=0.8]{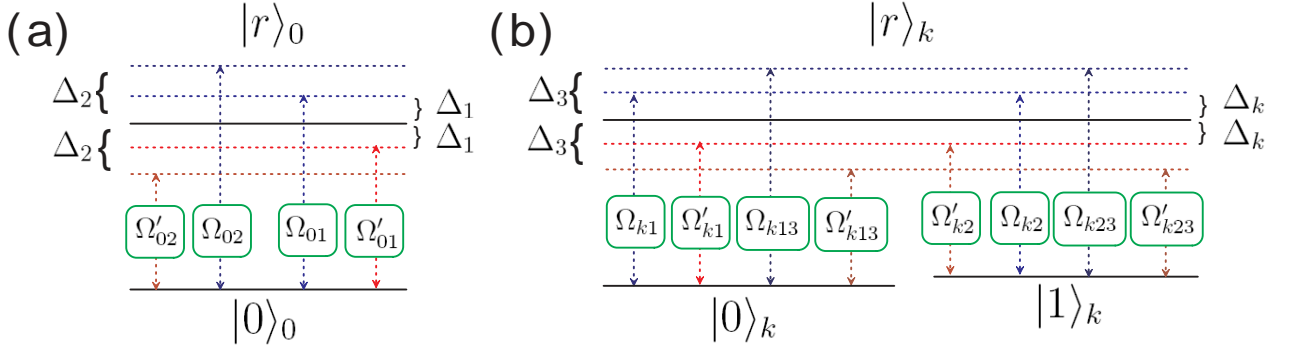}} \caption{(a)
Level configuration of the auxiliary atom $0$. (b) Level
configuration of the computational atom $k$ ($k=1,2$).}\label{fig1}
\end{figure}

We consider a system containing three atoms 0, 1 and 2 whose level
configurations are shown in Fig.~\ref{fig1}. The atom 0 has a ground
level $|0\rangle_0$ and a Rydberg state $|r\rangle_0$, and is used
as an auxiliary qubit. The transition
$|0\rangle_0\leftrightarrow|r\rangle_0$ is driven by two pairs of
laser pulses. One pair possesses Rabi frequencies $\Omega_{01}(t)$
and $\Omega_{01}'(t)$ with different detunings $\pm\Delta_1$,
respectively [positive (negative) corresponding to blue (red)
detuning] \cite{note1}. The other pair possesses Rabi frequencies
$\Omega_{02}(t)$ and $\Omega_{02}'(t)$ with different detunings
$\pm\Delta_2$, respectively. Besides, the atom $k$ ($k=1,2$) is
employed as a computational atom, possessing two ground levels
$|0\rangle_k$ and $|1\rangle_k$ as computational basis and a Rydberg
state $|r\rangle_k$ as an auxiliary level. The transition
$|j-1\rangle_k\leftrightarrow|r\rangle_k$ ($j=1,2$) is driven by two
pairs of laser pulses, one pair with Rabi frequencies
$\Omega_{kj}(t)$ and $\Omega_{kj}'(t)$ and different detunings
$\pm\Delta_k$, the other pair with Rabi frequencies
$\Omega_{kj3}(t)$ and $\Omega_{kj3}'(t)$ and different detunings
$\pm\Delta_3$. We assume that, between each two atoms, there exists
a Rydberg interaction with strength $V$ \cite{note2}.

\begin{figure}
\scalebox{0.6}{\includegraphics[scale=0.64]{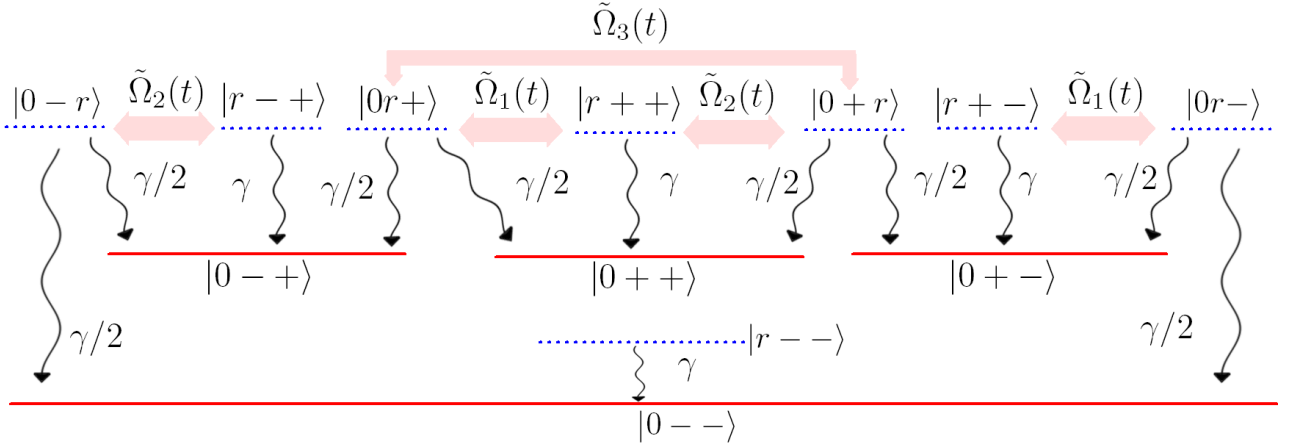}} \caption{The
subspaces of unitary evolution (levels with blue-dotted lines) and
dissipative evolution (levels with blue-dotted and red-solid
lines).}\label{fig2}
\end{figure}

To parameterize the Rabi frequencies for the implementation of the
NHQC, we set
\begin{eqnarray}\label{fp00}
&&\Omega_{k0}=\Omega_k(t)\cos(\theta_k/2),\
\Omega_{k0}'=\Omega_k'(t)\cos(\theta_k/2),\cr\cr
&&\Omega_{k1}=\Omega_k(t)\sin(\theta_k/2)e^{i\varphi_k},\
\Omega_{k1}'=\Omega_k'(t)\sin(\theta_k/2)e^{i\varphi_k},\cr\cr
&&\Omega_{k03}=\Omega_{k3}(t)\cos(\theta_k/2),\
\Omega_{k03}'=\Omega_{k3}'(t)\cos(\theta_k/2),\cr\cr
&&\Omega_{k13}=\Omega_{k3}(t)\sin(\theta_k/2)e^{i\varphi_k},\
\Omega_{k13}'=\Omega_{k3}'(t)\sin(\theta_k/2)e^{i\varphi_k},
\end{eqnarray}
with $\theta_k$ and $\varphi_k$ being two time-independent
parameters. In addition, we assume
\begin{eqnarray}\label{fp01}
&&\Omega_k(t)=e^{i\mu_k(t)}\bar{\Omega}_k(t),\
\Omega_k'(t)=e^{i\mu_k'(t)}\bar{\Omega}_k(t),\
\Omega_{0k}(t)=\Omega_{0k}'(t)=-\bar{\Omega}_{0k}(t),\cr\cr
&&\Omega_{13}(t)=e^{i\mu_{3}(t)}\bar{\Omega}_{k3}(t),\
\Omega_{13}'(t)=e^{i\mu_{3}'(t)}\bar{\Omega}_{k3}(t),\
\Omega_{23}(t)=\Omega_{23}'(t)=-\bar{\Omega}_{23}(t),
\end{eqnarray}
with
$\{\bar{\Omega}_k(t),\bar{\Omega}_{0k}(t),\bar{\Omega}_{k3}(t)\}$
being real functions. In the regime of Rydberg blockade, effective
Hamiltonian of the system can be derived as (detailed derivations
are shown in Appendix A)
\begin{eqnarray}\label{e4}
H_e(t)&=&H_{e_0}(t)+H_{e_1}(t)+H_{e_2}(t)+H_{e_3}(t),\cr\cr
H_{e_0}(t)&=&\tilde{\Omega}_{1}(t)e^{i\mu_1(t)}|r++\rangle_{012}\langle0r+|
+\tilde{\Omega}_{2}(t)e^{i\mu_2(t)}|r++\rangle_{012}\langle0+r|+\mathrm{H.c.}\cr\cr
H_{e_1}(t)&=&\tilde{\Omega}_{1}(t)e^{i\mu_1(t)}|r+-\rangle_{012}\langle
0r-|+\mathrm{H.c.},\cr\cr
H_{e_2}(t)&=&\tilde{\Omega}_{2}(t)e^{i\mu_2(t)}|r-+\rangle_{012}\langle
0-r|+\mathrm{H.c.},\cr\cr
H_{e_3}(t)&=&\tilde{\Omega}_{3}(t)e^{i\mu_3(t)}|0r+\rangle_{012}\langle0+r|+\mathrm{H.c.}
\end{eqnarray}
with
\begin{eqnarray}\label{e7}
&&\tilde{\Omega}_{k}(t)=\frac{2\bar{\Omega}_k(t)\bar{\Omega}_{0k}(t)}{\Delta_k},\
\tilde{\Omega}_{3}(t)=\frac{2\bar{\Omega}_{13}(t)\bar{\Omega}_{23}(t)}{\Delta_3},\cr\cr&&
|+\rangle_k=\cos(\theta_k/2)|0\rangle_k+\sin(\theta_k/2)e^{i\varphi_k}|1\rangle_k,\cr\cr&&
|-\rangle_k=\sin(\theta_k/2)|0\rangle_k-\cos(\theta_k/2)e^{i\varphi_k}|1\rangle_k.
\end{eqnarray}
According to Eq.~(\ref{e4}), when the auxiliary atom is initially
prepared in Rydberg state $|r\rangle_0$, and the computational atoms
are in arbitrary state in superposition of computational basis
$|\pm\rangle_k$, the evolution is restricted in an eight-dimensional
subspace
$\mathcal{B}=\mathrm{span}\{|r\pm\pm\rangle_{012},|0r\pm\rangle_{012},|0\pm
r\rangle_{012}\}$, which are shown by the levels with blue-dotted
lines in Fig.~\ref{fig2}.

\subsection{Evolution of the system in dissipative environment}

We now take dissipation into account, and analyze the evolution of
the system under the influence of dissipation. For computational
atom $k$, there exists two dissipative paths as
$|r\rangle_k\rightarrow|0\rangle_k$ and
$|r\rangle_k\rightarrow|1\rangle_k$. Considering the same decay rate
$\gamma/2$ for each path, the dissipation of atom $k$ can be
described by two Lindblad operators as $L_{\pm
k}=\sqrt{\gamma/2}|(1\pm1)/2\rangle_k\langle r|$. For the auxiliary
atom $0$, there exists a dissipative path
$|r\rangle_0\rightarrow|0\rangle_0$. Assuming the decay rate of
$|r\rangle_0\rightarrow|0\rangle_0$ is $\gamma$, dissipation of atom
can be described by the Lindblad operator
$L_0=\sqrt{\gamma}|0\rangle_0\langle r|$. The evolution of the total
system under the influence of dissipation is governed by the master
equation
\begin{eqnarray}\label{de1}
\dot{\rho}(t)=-i[H_e(t),\rho(t)]+\sum\limits_{\iota=-2}^{2}[L_\iota\rho(t)
L_\iota^{\dag}-\frac{1}{2}L_\iota^{\dag}L_\iota\rho(t)-\frac{1}{2}\rho(t)
L_\iota^{\dag}L_\iota],
\end{eqnarray}
with $\rho(t)$ being the density operator of the system. When
dissipation is taken into account, the evolution of the system is no
longer remained in the eight-dimensional subspace $\mathcal{B}$, but
confined to a twelve-dimensional subspace
$\mathcal{B}'=\mathrm{span}\{|r\pm\pm\rangle_{012},|0r\pm\rangle_{012},|0\pm
r\rangle_{012},|0\pm\pm\rangle_{012}\}$, which are shown by levels
with both blue-dotted and red-solid lines in Fig.~\ref{fig2}. The
nonzero matrix elements of $\dot{\rho}(t)$ is shown in Appendix B,
where the results indicate the density operator can be written as
\begin{eqnarray}\label{de2}
\rho(t)=e^{-\gamma t}U_e(t)\rho(0)U_e^\dag(t)\oplus\rho'(t),
\end{eqnarray}
with $U_e(t)$ being the evolution operator given by the equation
$i\dot{U}_e(t)=H_e(t)U_e(t)$, and $\rho'(t)$ being part of density
operator shown in Appendix B. We assume that the target operation is
$U_e^{\mathrm{ideal}}(T)=|r\rangle_0\langle r|\otimes U_{12}
+|0\rangle_0\langle0|\otimes\mathbbold{1}_{12}$
($\mathbbold{1}_{12}$ denotes the identity operation for atoms 1 and
2). Then, $U_e(T)\rho(0)U_e^\dag(T)$ in the ideal case should be
$|r\rangle_0\langle r|\otimes U_{12}\rho_{12}(0)U_{12}^\dag$, as the
auxiliary atom is initially prepared in the Rydberg state. On the
other hand, we can also see in Appendix B that all nonzero elements
of $\rho'(t)$ do not contain $|r\rangle_0$ component. Therefore, if
one measures the state of the auxiliary atom at $t=T$ and obtain the
result $|r\rangle_0$, we confirm that decays of atoms are not
happened, and the unitary operation $U_{12}$ on computational atoms
1 and 2 is successfully implemented. According to Eq.~(\ref{de2}),
the successful probability is $P_s=e^{-\gamma T}$. If we consider
the decay rate as $\gamma=1$kHz \cite{SSLPRA96,SXQPRA96}, the
successful probability is higher than 99\% when $T\leq10.1\mu$s, and
is still higher than 90\% when $T\leq105.4\mu$s, theoretically.

\section{Arbitrary single-qubit gates}

\subsection{Construction of evolution path with invariant-based reverse engineering}

In this section, let us firstly consider the implementation of
arbitrary single-qubit gates by using the computational atom 1 and
the auxiliary atom 0 with invariant-based reverse engineering. In
this case, laser pulses $\Omega_2(t)$, $\Omega_2'(t)$,
$\Omega_{02}(t)$, $\Omega_{02}'(t)$, $\Omega_{k3}(t)$ and
$\Omega_{k3}'(t)$ are switched off. According to the effective
Hamiltonian in Eq.~(\ref{e4}), the Hamiltonian for the
implementation of singlet-qubit gates read
\begin{eqnarray}\label{e8}
H_s(t)=\tilde{\Omega}_{1}(t)e^{i\mu_1(t)}|r+\rangle_{01}\langle0r|+\mathrm{H.c.}
=\tilde{\Omega}_{1}(t)\cos[\mu_1(t)]\sigma_x+\tilde{\Omega}_{1}(t)\sin[\mu_1(t)]\sigma_y+0\times\sigma_z,
\end{eqnarray}
with
\begin{eqnarray}\label{e9}
\sigma_x=|0r\rangle_{01}\langle r+|+\mathrm{H.c.},\
\sigma_y=-i|0r\rangle_{01}\langle r+|+\mathrm{H.c.}, \
\sigma_z=|0r\rangle_{01}\langle0r|-|r+\rangle_{01}\langle r+|,
\end{eqnarray}
which can be considered as generators of su(2) Lie algebra
satisfying commutation relations
\begin{eqnarray}\label{e10}
[\sigma_x,\sigma_y]=2i\sigma_z,\ [\sigma_y,\sigma_z]=2i\sigma_x, \
[\sigma_z,\sigma_x]=2i\sigma_y.
\end{eqnarray}

To find a dynamic invariant with Lie algebra, we consider a dynamic
invariant in a superposition of all generators of su(2) algebra as
\cite{TorronteguiPRA89}
\begin{eqnarray}\label{e11}
I_s(t)=\lambda_x(t)\sigma_x+\lambda_y(t)\sigma_y+\lambda_z(t)\sigma_z.
\end{eqnarray}
By substituting Eq.~(\ref{e11}) into Eq.~(\ref{lr1}), we obtain the
equations as
\begin{eqnarray}\label{e12}
&&\dot{\lambda}_x(t)=2\Omega_y(t)\lambda_z(t),\ \
\dot{\lambda}_y(t)=-2\Omega_x(t)\lambda_z(t),\ \
\dot{\lambda}_z(t)=2\Omega_x(t)\lambda_y(t)-2\Omega_y(t)\lambda_x(t),
\end{eqnarray}
with $\Omega_x(t)=\tilde{\Omega}_{1}(t)\cos[\mu_1(t)]$ and
$\Omega_y(t)=\tilde{\Omega}_{1}(t)\sin[\mu_1(t)]$. From
Eq.~(\ref{e12}), we can derive a constraint equation for
coefficients of the dynamic invariant $I_s(t)$ as
$\lambda_x^2(t)+\lambda_y^2(t)+\lambda_z^2(t)=C^2$ with $C$ being a
real constant. Accordingly, for $C=1$, $\lambda_x(t)$,
$\lambda_y(t)$ and $\lambda_z(t)$ can be parameterized as
\begin{eqnarray}\label{e13}
\lambda_x(t)=\sin\beta_1\sin\beta_2,\ \
\lambda_y(t)=\sin\beta_1\cos\beta_2,\ \ \lambda_z(t)=\cos\beta_1,
\end{eqnarray}
with two time-dependent parameters $\beta_1$ and $\beta_2$. Using
Eqs.~(\ref{e12}-\ref{e13}), we can reversely solve $\Omega_x(t)$ and
$\Omega_y(t)$ as
\begin{eqnarray}\label{e14}
&&\Omega_x(t)=(\dot{\beta}_2\sin\beta_2\tan\beta_1-\dot{\beta_1}\cos\beta_2)/2,\
\
\Omega_y(t)=(\dot{\beta}_2\cos\beta_2\tan\beta_1+\dot{\beta_1}\sin\beta_2)/2.
\end{eqnarray}
In addition, eigenvectors of the dynamic invariant $I_s(t)$ can also
be obtained as
\begin{eqnarray}\label{e15}
|\phi_+^s(t)\rangle=\cos\frac{\beta_1}{2}|0r\rangle_{01}+ie^{-i\beta_2}\sin\frac{\beta_1}{2}|r+\rangle_{01},\
|\phi_-^s(t)\rangle=ie^{i\beta_2}\sin\frac{\beta_1}{2}|0r\rangle_{01}+\cos\frac{\beta_1}{2}|r+\rangle_{01},\
\end{eqnarray}
whose eigenvalues are 1 and -1, respectively. By applying the
results in Eqs.~(\ref{lr8}-\ref{lr9}), the time derivatives of
dynamic phases and geometric phases acquired by
$|\phi_\pm^s(t)\rangle$ can be respectively calculated by
\begin{eqnarray}\label{e16}
\dot{\vartheta}_\pm(t)=\mp\frac{\dot{\beta}_2\sin^2\beta_1}{2\cos\beta_1},\
\ \dot{\Theta}_\pm(t)=\pm\dot{\beta}_2\sin^2\frac{\beta_1}{2}.
\end{eqnarray}

In the implementation of the single qubit gate, the computational
subspace is spanned by $|r\pm\rangle_{01}$. The vector
$|r-\rangle_{01}$ is dynamically decoupled to the Hamiltonian
$H_s(t)$. Therefore, the evolution with initial state
$|r+\rangle_{01}$ is mainly studied here. We consider the boundary
condition as $\beta_1(0)=\beta_1(T)=0$ for cyclic evolution of
eigenvectors as $|\phi_\pm^s(0)\rangle=|\phi_\pm^s(T)\rangle$. In
this case, the boundary condition of $\beta_2(t)$ is irrelevant to
the cyclic evolution condition according to Eq.~(\ref{e15}). In
addition, the evolution with initial state $|r+\rangle_{01}$ is
along the eigenvector $|\phi_-^s(t)\rangle$. In order to eliminate
the dynamic phase $\vartheta_-(T)$ and obtain a pure geometric phase
$\Theta_-(T)=\Theta_s$, we design parameters $\beta_1(t)$ and
$\beta_2(t)$ by dividing the time interval $[0,T]$ into three parts
$[0,\tau_1]$, $[\tau_1,\tau_2]$ and $[\tau_2,T]$. When
$t\in[0,\tau_1]$, $\beta_1(t)$ increases from 0 to $\pi$, and
$\beta_2(t)$ remains an undetermined parameter. When
$t\in[\tau_1,\tau_2]$, we keep $\beta_1(t)=\pi$, and let
$\beta_2(t)=\beta_2(\tau_1)-\Theta_s(t-\tau_1)/(\tau_2-\tau_1)$.
When $t\in[\tau_2,T]$, we set
$\beta_1(t)=\beta_1[\tau_1(T-t)/(T-\tau_2)]$,
$\beta_2(t)=-\Theta_s+\beta_2[\tau_1(T-t)/(T-\tau_2)]$, which meet
the boundary conditions $\beta_1(\tau_2)=\beta_1(\tau_1)=\pi$,
$\beta_1(T)=\beta_1(0)=0$,
$\beta_2(\tau_2)=-\Theta_s+\beta_2(\tau_1)$, and
$\beta_2(T)=-\Theta_s$. With above assumptions of $\beta_1(t)$ and
$\beta_2(t)$, the dynamic (geometric) phase acquired in the time
interval $[0,\tau_1]$ is nullified by that acquired in time interval
$[\tau_2,T]$ (see Appendix C for details). Moreover, in the time
interval $[\tau_1,\tau_2]$, as $\beta_1(t)$ is kept at $\pi$, the
dynamic phase acquired is zero according to Eq.~(\ref{e16}). On the
other hand, the geometric phase acquired is
\begin{eqnarray}\label{e17}
\Theta_-(\tau_2)-\Theta_-(\tau_1)=-\int_{\tau_1}^{\tau_2}\dot{\beta}_2(t)\sin^2[\frac{\beta_1(t)}{2}]dt
=\beta_2(\tau_1)-\beta_2(\tau_2)=\Theta_s.
\end{eqnarray}
Consequently, the total dynamic phase and the geometric phase
acquired in $[0,T]$ are $\vartheta_-(T)=0$ and
$\Theta_-(T)=\Theta_s$. Therefore, the evolution operator can be
described as
\begin{eqnarray}\label{e18}
U_s(T,0)=e^{i\Theta_s}|r+\rangle_{01}\langle
r+|+|r-\rangle_{01}\langle r-|=e^{i\Theta_s/2}|r\rangle_0\langle
r|\otimes e^{i\Theta_s\vec{n}_1\cdot\vec{\sigma_1}/2},
\end{eqnarray}
with
$\vec{n}_1=[\sin\theta_1\cos\varphi_1,\sin\theta_1\sin\varphi_1,\cos\theta_1]$,
$\vec{\sigma_1}=[\sigma_{x1},\sigma_{y1},\sigma_{z1}]$,
$\sigma_{x1}=|0\rangle_1\langle1|+\mathrm{H.c.}$,
$\sigma_{y1}=-i|0\rangle_1\langle1|+\mathrm{H.c.}$, and
$\sigma_{z1}=|0\rangle_1\langle0|-|1\rangle_1\langle1|$. Up to a
global phase $\Theta_s/2$, Eq.~(\ref{e18}) represents a rotation
operator around the axis $\vec{n}$ with rotation angle $\Theta_s/2$,
which can generate arbitrary single-qubit gates
\cite{JLNPRA100,LZTPRA93}. Specially, when
$(\Theta_s,\theta_1,\varphi_1)=\pi(1,1/2,1)$, we get a Not gate for
atom 1 as $U_N=\sigma_{x1}$; when $\theta_1=\pi$, we obtain a
$\Theta_s$-phase gate for atom 1 as
$U_{\Theta_s}=\mathrm{diag}[1,\exp(i\Theta_s)]$; when
$(\Theta_s,\theta_1,\varphi_1)=\pi(1,-1/4,0)$, a Hadamard gate
$U_H=(\sigma_{x1}+\sigma_{z1})/\sqrt{2}$ for atom 1 is realized.

The design of parameters $\beta_1(t)$ and $\beta_2(t)$ can be
further simplified. In fact, in time interval $[\tau_1,\tau_2]$,
$\beta_1(t)=\pi$, we have $\Omega_x(t)=\Omega_y(t)=0$ according to
Eq.~(\ref{e14}). As a result, the system does not evolve in time
interval $[\tau_1,\tau_2]$ as $H_s(t)=0$. Therefore, the time
interval $[\tau_1,\tau_2]$ can be reduced by letting
$\tau_2\rightarrow\tau_1$. In this case, we also have the change of
geometric phase as
\begin{eqnarray}\label{e171}
\Delta\Theta_-=\lim_{\tau_2\rightarrow\tau_1}[\Theta_-(\tau_2)-\Theta_-(\tau_1)]
=\lim_{\tau_2\rightarrow\tau_1}\int_{\tau_1}^{\tau_2}\frac{\Theta_sdt}{\tau_2-\tau_1}
=\Theta_s.
\end{eqnarray}
Therefore, in the limit of $\tau_2=\tau_1=\tau$, we only need to
increase $\beta_1(t)$ from 0 to $\pi$ in time interval $[0,\tau]$
and set $\beta_1(t)=\beta_1[\tau(T-t)/(T-\tau)]$,
$\beta_2(t)=-\Theta_s+\beta_2[\tau(T-t)/(T-\tau)]$ in time interval
$[\tau,T]$. Besides, as $\Omega_x(\tau)=\Omega_y(\tau)=0$, the
expressions of $\Omega_x(t)$ and $\Omega_y(t)$ can be still
continuous functions if we make the value of $\beta_2$ increased by
$-\Theta_s$ at the moment of $t=\tau$.

\subsection{Selections of parameters for robust control}

In Sec. IV, we have determined boundary conditions for parameters
$\beta_1(t)$ and $\beta_2(t)$. However, specific expressions for
$\beta_1(t)$ and $\beta_2(t)$ are still not given. In a real
implementation of atomic quantum gates by using laser pulses,
systematic errors of laser pulses due to imperfections of devices
may be troublesome factors to obtain high gate fidelities. In order
to realize robust control, let us now discuss selections of
parameters $\beta_1(t)$ and $\beta_2(t)$ with the help of optimal
method \cite{RuschhauptNJP14} by nullifying systematic error
sensitivity. In presence of the systematic error with error
coefficient $\epsilon$, the Rabi frequencies of laser pulses become
$\Omega_{01}(t)\rightarrow(1+\epsilon)\Omega_{01}(t)$,
$\Omega_{1}(t)\rightarrow(1+\epsilon)\Omega_{1}(t)$. According to
Eq.~(\ref{e7}), we have
$\tilde{\Omega}_{1}(t)\rightarrow[2(1+\epsilon)^2\bar{\Omega}_{0}(t)\bar{\Omega}_{01}(t)]/\Delta_1
=(1+2\epsilon)\tilde{\Omega}_{1}(t)+\mathcal{O}(\epsilon^2)$, with
$\mathcal{O}(\epsilon^2)$ being the terms with orders equal or
higher than $\epsilon^2$. Thus, in the effective Hamiltonian, the
effective error coefficient is $\tilde{\epsilon}=2\epsilon$, and the
erroneous effective Hamiltonian is
$\tilde{H}_s(t)=(1+\tilde{\epsilon})H_s(t)$. With the help of
time-dependent perturbation theory, one can derive
\cite{RuschhauptNJP14,YXTPRA97}
\begin{eqnarray}\label{e182}
|\psi_s^{\tilde{\epsilon}}(T)\rangle=|\psi_s(T)\rangle
-i\tilde{\epsilon}\int_0^TdtU_s(T,t)H_s(t)|\psi_s(t)\rangle+\mathcal{O}(\tilde{\epsilon}^2),
\end{eqnarray}
where $|\psi_s(t)\rangle$ ($|\psi_s^{\tilde{\epsilon}}(t)\rangle$)
is the state of the system without (with) systematic errors. As the
state $|r-\rangle_{01}$ is dynamically decoupled to the Hamiltonian
$H_s(t)$, and the evolution with initial state $|r+\rangle_{01}$ is
described by $|\psi_s(t)\rangle=e^{i\alpha_-(t)}|\phi_-(t)\rangle$,
we estimate the fidelity as
\begin{eqnarray}\label{e19}
F_s=1-\tilde{\epsilon}^2|\int_0^Te^{2i\alpha_-(t)}\langle\phi_+(t)|H_s(t)|\phi_-(t)\rangle
dt|^2+\mathcal{O}(\tilde{\epsilon}^3),
\end{eqnarray}
with $\alpha_+(t)=-\alpha_-(t)$ being considered. Therefore, the
systematic error sensitivity $Q_s$ can be calculated as
\begin{eqnarray}\label{e20}
Q_s=-\frac{\partial^2F_s}{2\partial\tilde{\epsilon}^2}
=|\int_0^Te^{i\chi(t)}\dot{\beta}_1\sin^2\beta_1dt|^2,
\end{eqnarray}
with $\chi(t)=\beta_2(t)+2\alpha_-(t)$. To nullify $Q_s$, we
consider $\chi(t)=\chi_0\{2\beta_1(t)-2\sin[2\beta_1(t)]\}$
($t\in[0,\tau)$) \cite{RuschhauptNJP14,DaemsPRL111}, with $\chi_0$
being a time-independent parameter. Noticing at $t=\tau$, $\chi(t)$
have a shift as $\Delta\chi=\Delta\beta_2+2\Delta\Theta_-=\Theta_s$,
one can derive
$\chi(t)=\Theta_s+\chi_0\{2\beta_1(t)-\sin[2\beta_1(t)]\}$
($t\in[\tau,T]$) and
$Q_s=\sin^2(\chi_0\pi)\sin^2(\Theta_s/2)/\chi_0^2$ (see Fig.~3(a)).
Therefore, $Q_s$ can be nullified when $\chi_0$ is a nonzero integer
[$Q_s\rightarrow\pi^2\sin^2(\Theta_s/2)$, $(\chi_0\rightarrow0)$].
Considering the fact that a larger value of $\chi_0$ leads a longer
operation time when the maximal pulse intensity is fixed, we set
$\chi_0=1$. In this case, we have $\beta_2(t)=4\sin^3[\beta_1(t)]/3$
with $t\in[0,\tau)$ and $\beta_2(t)=-\Theta_s+4\sin^3[\beta_1(t)]/3$
with $t\in[\tau,T]$. To make the pulses continuous and vanish at
boundary, we consider $\beta_1(t)$ as $\beta_1(t)=\pi\sin^2(\pi
t/T)$ with $\tau=T/2$ being set.

\begin{figure}
\scalebox{0.6}{\includegraphics[scale=0.7]{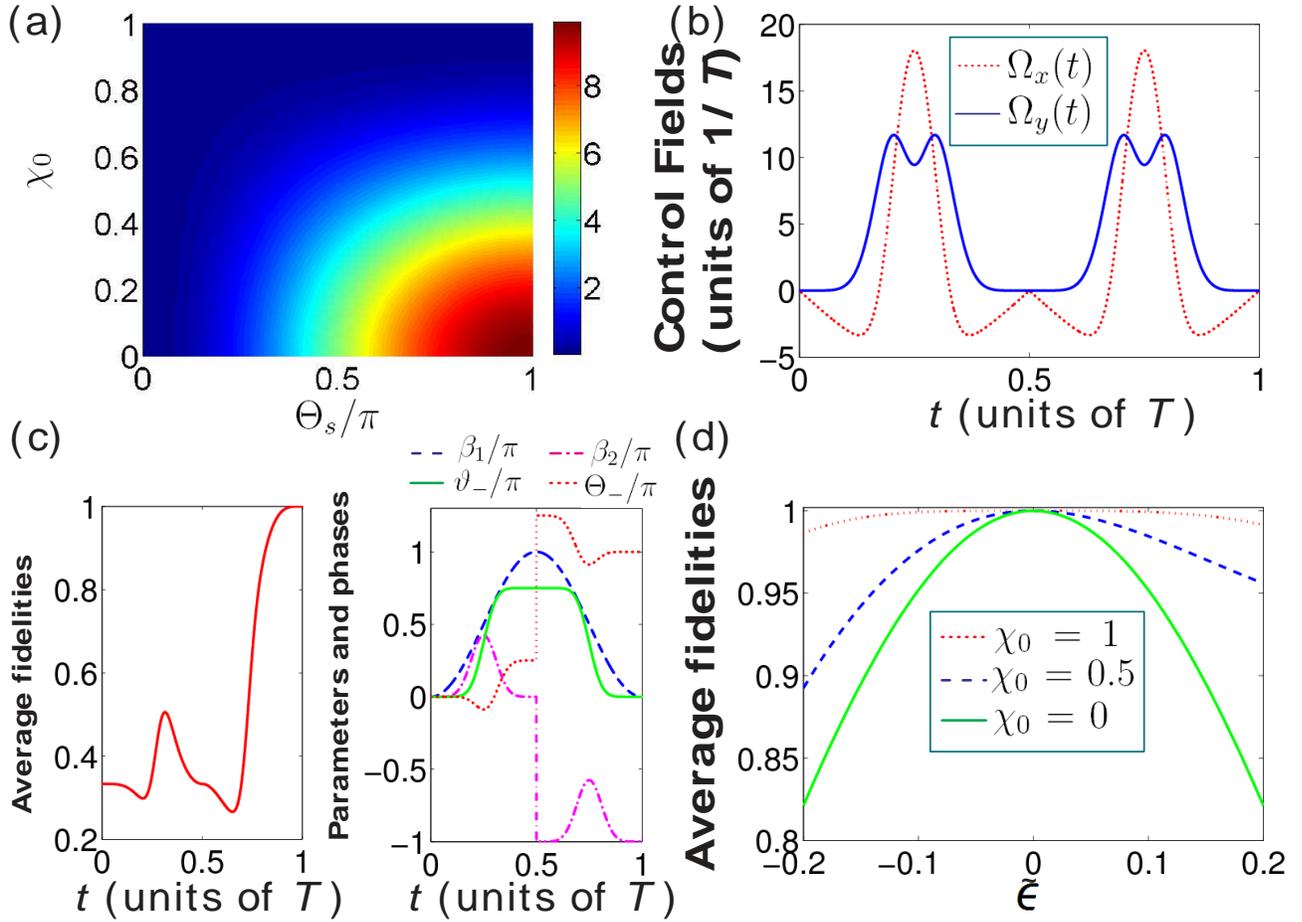}} \caption{(a)
$Q_s$ versus $\Theta_s/\pi$ and $\chi_0$. (b) $\Omega_x(t)$ and
$\Omega_y(t)$ versus $t$. (c) Average fidelities $\bar{F}_N(t)$,
parameters $\beta_1(t)$, $\beta_2(t)$, and phases $\vartheta_-(t)$,
$\Theta_-(t)$ in the implementation of the Not gate versus $t$ with
the effective Hamiltonian. (d) The final average fidelities
$\bar{F}_N(T)$ versus $\tilde{\epsilon}$ with the effective
Hamiltonian and parameters $\chi_0=1$ (red-dotted line),
$\chi_0=0.5$ (blue-dashed line) and $\chi_0=0$ (green-solid
line).}\label{fig3}
\end{figure}

With the parameters designed above, we plot $\Omega_x(t)$ and
$\Omega_y(t)$ versus $t$ in Fig.~\ref{fig3}(b), from which we can
further obtain
$\tilde{\Omega}_{\max}^s=\max\limits_{t\in[0,T]}[\tilde{\Omega}_1(t)]=20.35/T$.
Moreover, we plot the average fidelity (see Appendix D for details)
of the implementation of the Not gate versus $t$ in
Fig.~\ref{fig3}(c) as example to show the validity of the reverse
engineering and the parameter selections discussed above. The
average fidelity is defined as \cite{ZanardiPRA70,PedersenPLA367}
\begin{eqnarray}\label{e22}
\bar{F}_N(t)=\frac{1}{\mathcal{N}(\mathcal{N}+1)}\{\mathrm{Tr}[M(t)M^\dag(t)]
+|\mathrm{Tr}[M(t)]|^2\},
\end{eqnarray}
with $M(t)=\mathcal{P}_c\tilde{U}^\dag_NU_s(t)\mathcal{P}_c$,
$\tilde{U}_N=|0r\rangle_{01}\langle0r|+|r\rangle_0\langle r|\otimes
U_N$, $\mathcal{P}_c=|r0\rangle_{01}\langle
r0|+|r1\rangle_{01}\langle r1|$ being the projection operator onto
the computational subspace, and $\mathcal{N}=2$ for the
two-dimensional computational subspace. As shown in
Fig.~\ref{fig3}(c), in accordance with the expectation, the average
fidelity become unity at $t=T$. In addition, we also plot the
variations of parameters $\beta_1(t)$, $\beta_2(t)$ and acquired
dynamic phases $\vartheta_-(t)$ and geometric phase $\Theta_-(t)$ in
Fig.~\ref{fig3}(c), where we can find the dynamic phase
$\vartheta_-(t)$ finally vanishes at $t=T$, while the geometric
phase reaches the preset value $\Theta_-(T)=\Theta_s=\pi$. This
means we get pure geometric phase in the process. Therefore, the
reverse engineering and the parameter selections are effectively
applied on the effective Hamiltonian $H_s(t)$ for the implementation
of single qubit gates.

On the other hand, we plot the average fidelities $\bar{F}_N(T)$ at
the final time $T$ versus $\tilde{\epsilon}$ with $\chi_0=1$,
$\chi_0=0.5$ and $\chi_0=0$ in Fig.~\ref{fig3}(d) to show the
robustness against systematic errors. According to the red-dotted
line in Fig.~\ref{fig3}(d), $\bar{F}_N(T)$ keeps higher than 0.9864
when $\tilde{\epsilon}\in[-0.2,0.2]$ with $\chi_0=1$. This result
shows that the implementation of the Not gate with the optimal
parameter $\chi_0=1$ is quite insensitive to the systematic errors.
We also see from the green-solid line in Fig.~\ref{fig3}(d) that the
average fidelity falls to 0.8212 when $\tilde{\epsilon}=\pm0.2$ with
$\chi_0=0$. Noticing that when $\chi_0=0$, the implementation of the
single qubit gate become the same as that in the conventional NHQC
with only $\sigma_x$ control, the protocol can improve the
robustness against systematic errors. Moreover, see from the
blue-dashed line in Fig.~\ref{fig3}(d), the robustness against
systematic errors with $\chi_0=0.5$ is between $\chi_0=0$ and
$\chi_0=1$. As the total operation time increases with $\chi_0$ when
the maximal intensity of laser pulses is fixed, and the successful
probability of the protocol deceases with the increase of total
operation time as shown in Sec. IIIB, one may adjust the value of
$\chi_0$ between 0 and 1 to make trade-off between the robustness
against systematic errors and dissipation.

\section{Two-qubit entangling gates}

In this section, let us further study the implementations of
two-qubit entangling gates. Based on the effective Hamiltonian
$H_e(t)$ in Eq.~(\ref{e4}), we consider the conditions
$|\Omega_{e3}(t)|\gg\{|\Omega_{e1}(t)|,|\Omega_{e2}(t)|\}$,
$\mu_3(t)=0$ and $\dot{\tilde{\Omega}}_3(t)=0$. $H_{e_3}$ can be
diagonalized as
\begin{eqnarray}\label{e23}
H_{e_3}=\tilde{\Omega}_{3}(|\Phi_+\rangle\langle\Phi_+|-|\Phi_-\rangle\langle\Phi_-|),
\end{eqnarray}
with
$|\Phi_\pm\rangle=(|0r+\rangle_{012}\pm|0+r\rangle_{012})/\sqrt{2}$.
By performing a rotation transform with $\exp(-iH_{e_3}t)$, the
effective Hamiltonian can be transformed into
\begin{eqnarray}\label{e24}
\bar{H}_e(t)&=&\bar{H}_{e_0}(t)+\bar{H}_{e_1}(t)+\bar{H}_{e_2}(t),\cr\cr
\bar{H}_{e_0}(t)&=&\frac{\Omega_{e_1}(t)}{\sqrt{2}}|r++\rangle_{012}(\langle\Phi_+|e^{-i\tilde{\Omega}_3t}+\langle\Phi_-|e^{i\tilde{\Omega}_3t})\cr\cr
&+&\frac{\Omega_{e_2}(t)}{\sqrt{2}}|r++\rangle_{012}(\langle\Phi_+|e^{-i\tilde{\Omega}_3t}-\langle\Phi_-|e^{i\tilde{\Omega}_3t})+\mathrm{H.c.}\cr\cr
\bar{H}_{e_1}(t)&=&\Omega_{e_1}(t)|r+-\rangle_{012}\langle
0r-|+\mathrm{H.c.},\cr\cr
\bar{H}_{e_2}(t)&=&\Omega_{e_2}(t)|r-+\rangle_{012}\langle
0-r|+\mathrm{H.c.}
\end{eqnarray}
After omitting terms with high oscillation frequencies
$\pm\tilde{\Omega}_3$, the effective Hamiltonian can be simplified
as
\begin{eqnarray}\label{e25}
\bar{H}_e'(t)&=&\bar{H}_{e_1}(t)+\bar{H}_{e_2}(t),\cr\cr
\bar{H}_{e_1}(t)&=&\Omega_{e_1}(t)|r+-\rangle_{012}\langle
0r-|+\mathrm{H.c.},\cr\cr
\bar{H}_{e_2}(t)&=&\Omega_{e_2}(t)|r-+\rangle_{012}\langle
0-r|+\mathrm{H.c.}
\end{eqnarray}

According to Eq.~(\ref{e25}), evolutions in the subspaces
$\mathcal{S}_1=\mathrm{span}\{|r+-\rangle_{012},|0r-\rangle_{012}\}$
and
$\mathcal{S}_2=\mathrm{span}\{|r-+\rangle_{012},|0-r\rangle_{012}\}$
are independent, and the states $|r++\rangle_{012}$ and
$|r--\rangle_{012}$ are dynamically decoupled to the effective
Hamiltonian $\bar{H}_e'(t)$. Similar to Eq.~(\ref{e9}), we can make
the following definitions as
\begin{eqnarray}\label{fdp1}
&&\sigma_{x}^{(1)}=|0r-\rangle_{01}\langle r+-|+\mathrm{H.c.},\
\sigma_{y}^{(1)}=-i|0r-\rangle_{01}\langle r+-|+\mathrm{H.c.},
\cr\cr
&&\sigma_{z}^{(1)}=|0r-\rangle_{01}\langle0r|-|r+-\rangle_{01}\langle
r+-|,\ \sigma_{x}^{(2)}=|0-r\rangle_{01}\langle
r-+|+\mathrm{H.c.},\cr\cr
&&\sigma_{y}^{(2)}=-i|0-r\rangle_{01}\langle r-+|+\mathrm{H.c.}, \
\sigma_{z}^{(2)}=|0-r\rangle_{01}\langle0r|-|r-+\rangle_{01}\langle
r-+|,
\end{eqnarray}
which satisfying
\begin{eqnarray}\label{fdp2}
[\sigma_x^{(j)},\sigma_y^{(j)}]=2i\sigma_z^{(j)},\
[\sigma_y^{(j)},\sigma_z^{(j)}]=2i\sigma_x^{(j)}, \
[\sigma_z^{(j)},\sigma_x^{(j)}]=2i\sigma_y^{(j)},\
[\sigma_q^{(1)},\sigma_{q'}^{(2)}]=0\ (q,q'=x,y,z).\ \
\end{eqnarray}
Therefore, we can separately investigate evolutions in the subspaces
$\mathcal{S}_1$ and $\mathcal{S}_2$ with su(2) algebra. Similar to
the process in Sec. IVA, we can derive an invariant $I_2(t)$ as
\begin{eqnarray}\label{fdp3}
I_2(t)=\sum\limits_{j=1,2}\sum\limits_{q=x,y,z}\lambda_q^{(j)}(t)\sigma_q^{(j)},
\end{eqnarray}
with
\begin{eqnarray}\label{fdp4}
\lambda_x^{(j)}(t)=\sin\beta^{(j)}_1\sin\beta^{(j)}_2,\ \
\lambda_y^{(j)}(t)=\sin\beta^{(j)}_1\cos\beta^{(j)}_2,\ \
\lambda_z^{(j)}(t)=\cos\beta^{(j)}_1,
\end{eqnarray}
and eigenvectors of the dynamic invariant $I_2(t)$ are given by
\begin{eqnarray}\label{fdp5}
&&|\phi_+^{(1)}(t)\rangle=\cos\frac{\beta^{(1)}_1}{2}|0r-\rangle_{01}+ie^{-i\beta^{(1)}_2}\sin\frac{\beta^{(1)}_1}{2}|r+-\rangle_{01},\cr\cr
&&|\phi_+^{(2)}(t)\rangle=\cos\frac{\beta^{(2)}_1}{2}|0-r\rangle_{01}+ie^{-i\beta^{(2)}_2}\sin\frac{\beta^{(2)}_1}{2}|r-+\rangle_{01},\cr\cr
&&|\phi_-^{(1)}(t)\rangle=ie^{i\beta^{(1)}_2}\sin\frac{\beta^{(1)}_1}{2}|0r-\rangle_{01}+\cos\frac{\beta^{(1)}_1}{2}|r+-\rangle_{01},\cr\cr
&&|\phi_-^{(2)}(t)\rangle=ie^{i\beta^{(2)}_2}\sin\frac{\beta^{(2)}_1}{2}|0-r\rangle_{01}+\cos\frac{\beta^{(2)}_1}{2}|r-+\rangle_{01}.
\end{eqnarray}
In addition, the time derivatives of dynamic phases and geometric
phases acquired by $|\phi_\pm^{(j)}(t)\rangle$ can be respectively
calculated by
\begin{eqnarray}\label{fdp6}
\dot{\vartheta}_\pm^{(j)}(t)=\mp\frac{\dot{\beta}_2^{(j)}\sin^2\beta_1^{(j)}}{2\cos\beta_1^{(j)}},\
\
\dot{\Theta}_\pm^{(j)}(t)=\pm\dot{\beta}_2^{(j)}\sin^2\frac{\beta_1^{(j)}}{2}.
\end{eqnarray}

By applying the parameter design of $\beta_1(t)$ and $\beta_2(t)$ in
Sec. IV to parameters $\beta^{(j)}_1$ and $\beta^{(j)}_2$,
evolutions with initial states $|r+-\rangle_{012}$ and
$|r-+\rangle_{012}$ will move cycling along
$|\phi_-^{(1)}(t)\rangle$ and $|\phi_-^{(2)}(t)\rangle$ with
geometric phases $\bar{\Theta}_1$ and $\bar{\Theta}_2$ acquired
while the dynamic phases being eliminated. Then, the operation on
computational atoms 1 and 2 reads
\begin{eqnarray}\label{e26}
U_{12}(T,0)&=&|++\rangle_{12}\langle++|+e^{i\bar{\Theta}_1}|+-\rangle_{12}\langle+-|
+e^{i\bar{\Theta}_2}|-+\rangle_{12}\langle-+|+|--\rangle_{12}\langle--|\cr\cr
&=&e^{-i\bar{\Theta}_1/2}|+\rangle_1\langle+|\otimes
e^{-i\bar{\Theta}_1\vec{n}_2\cdot\vec{\sigma_2}/2}+e^{i\bar{\Theta}_2/2}|-\rangle_1\langle-|\otimes
e^{i\bar{\Theta}_2\vec{n}_2\cdot\vec{\sigma_2}/2},
\end{eqnarray}
with
$\vec{n}_2=[\sin\theta_2\cos\varphi_2,\sin\theta_2\sin\varphi_2,\cos\theta_2]$,
$\vec{\sigma_2}=[\sigma_{x2},\sigma_{y2},\sigma_{z2}]$,
$\sigma_{x2}=|0\rangle_2\langle1|+\mathrm{H.c.}$,
$\sigma_{y2}=-i|0\rangle_2\langle1|+\mathrm{H.c.}$, and
$\sigma_{z2}=|0\rangle_2\langle0|-|1\rangle_2\langle1|$.
Equation~(\ref{e26}) can be considered as a controlled
arbitrary-angle-rotation gate for atom 2 with atom 1 being the
control qubit. For example, when
$(\bar{\Theta}_1,\bar{\Theta}_2,\theta_1,\varphi_1,\theta_2,\varphi_2)=\pi(0,1,0,0,1/2,1)$,
the operation on atoms 1 and 2 is a controlled-Not (C-Not) gate
$U_{CN}=|0\rangle_1\langle0|\otimes\mathbbold{1}_2+|1\rangle_1\langle1|\otimes\sigma_{x2}$,
with $\mathbbold{1}_2$ being the identity operation for atom 2; when
$(\bar{\Theta}_1,\theta_1,\varphi_1,\theta_2)=\pi(0,0,0,1)$, the
controlled-$\bar{\Theta}_2$-phase gate of atoms 1 and 2 as
$U_{C\bar{\Theta}_2}=|0\rangle_1\langle0|\otimes\mathbbold{1}_2
+|1\rangle_1\langle1|\otimes\mathrm{diag}[1,e^{i\bar{\Theta}_2}]_2$
are realized.

\section{Numerical analysis and Discussions}

\subsection{Numerical analysis of single-qubit gate}

\begin{figure}
\scalebox{0.6}{\includegraphics[scale=0.7]{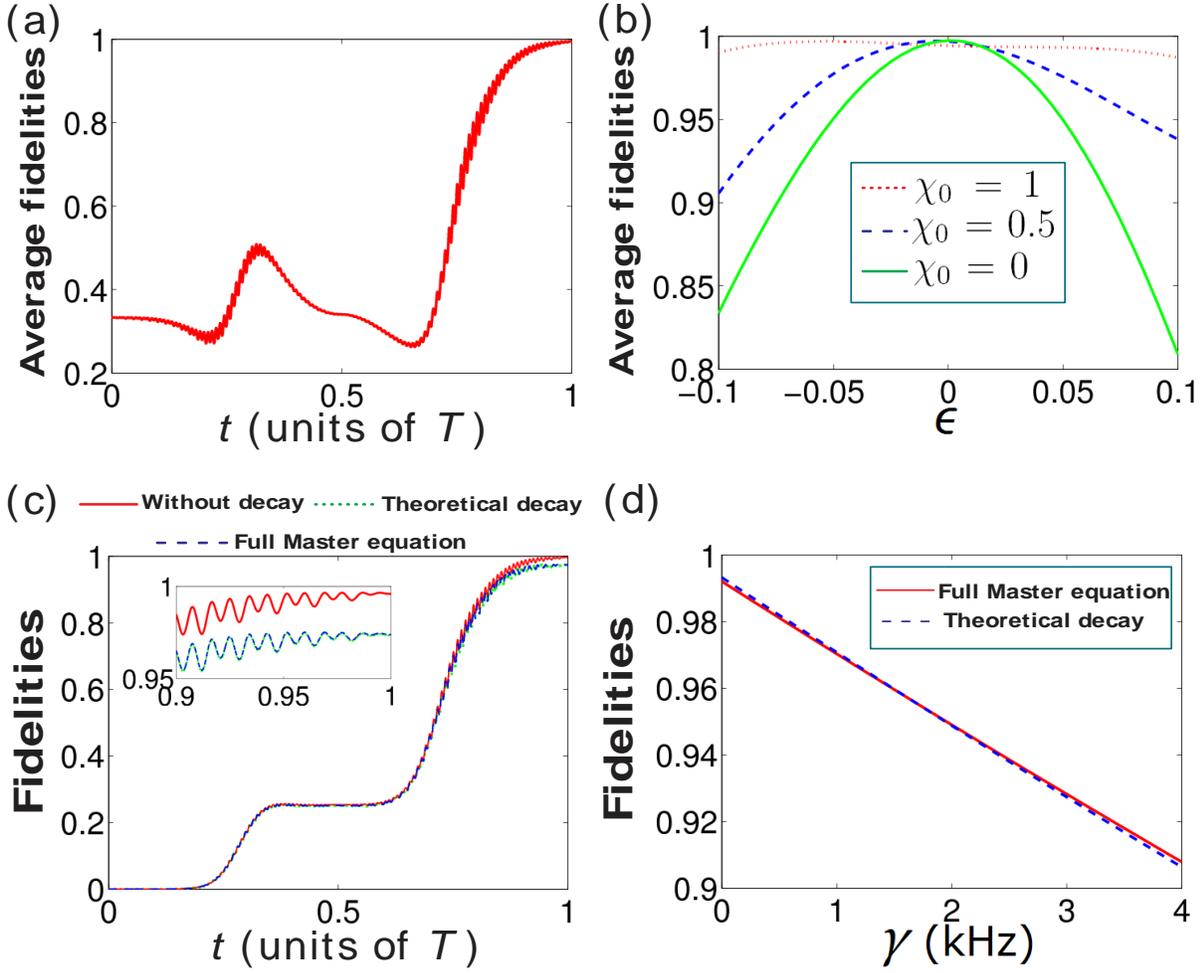}}
\caption{(a)Average fidelities $\bar{F}_N(t)$ of the implementation
of the Not gate versus $t$ with the full Hamiltonian. (b) The final
average fidelities $\bar{F}_N(T)$ versus $\epsilon$ with the full
Hamiltonian and parameters $\chi_0=1$ (red-dotted line),
$\chi_0=0.5$ (blue-dashed line) and $\chi_0=0$ (green-solid line).
(c) Fidelities $F_N(t)$ of the implementation of the Not gate versus
$t$ calculated by the master equation with initial state
$|r0\rangle_{01}$. Red-solid line: $\gamma=0$, green-dotted line:
$\gamma=1$kHz with theoretical result in Eq.~(\ref{de2}),
blue-dashed line: $\gamma=1$kHz with the full master equation. (d)
Final fidelities $F_N(T)$ of the implementation of the Not gate
versus decay rate $\gamma$ with initial state $|r0\rangle_{01}$ by
using theoretical result in Eq.~(\ref{de2}) (blue-dashed line) and
the full master equation (red-solid line).}\label{fig4}
\end{figure}

Let us now analyze the evolution based on the full Hamiltonian
$H(t)$ in Eq.~(\ref{fp1}). Firstly, we check the performance of
single-qubit gate by taking the Not gate as an example. To make the
Rydberg blockade and effective Hamiltonian valid, we choose
$V=7200/T$ and $\Delta_1=360/T$. Considering a reported Rydberg
interaction strength $V=2\pi\times50$MHz
\cite{WilkPRL104,LDXPRA98,ZRHPRA101}, where the center-to-center
distance between Rydberg atoms is about $d=3.755\mu$m with van der
Waals coefficient $C^{s}_6(r,r)=8.8\times10^{11}\mu\mathrm{m}^6/$s
\cite{LYMPRA91,SingerJPB38}, the total interaction time is
$T=22.9\mu$s, which gives the successful probability as
$P_s=99.74\%$ for probing state $|r\rangle_0$ of the auxiliary atom.
Furthermore, although the Stark shift
$-2V(\bar{\Omega}_1^2|r+\rangle_{01}\langle
r+|+\bar{\Omega}_{01}^2|0r\rangle_{01}\langle0r|)/(V^2-\Delta_1^2)$
due to the second order perturbation can be neglected when
$V\gg\Delta_1$, we can still eliminate the Stark shift by using
auxiliary pulses or levels \cite{SSLPRA96,LDXPRA98} to loose the
requirement on the ratio between $V$ and $\Delta_1$. We plot the
average fidelity $\bar{F}_N(t)$ versus $t$ with the full Hamiltonian
in Fig.~\ref{fig4}(a). From Fig.~\ref{fig4}(a), up to small
oscillations, the curve matches well with that plotted in
Fig.~\ref{fig3}(c) with the effective Hamiltonian. The final average
fidelity $\bar{F}_N(T)$ is 0.9943. Thus, the considered Rydberg
interaction strength $V$ and the detuning $\Delta_1$ are proper for
the construction of the effective Hamiltonian. In general, one may
increase the ratios $\tilde{\Omega}_{\max}^s/\Delta_1$ and
$\Delta_1/V$ to further improve the fidelity. However, the total
interaction time would also increase, and the successful probability
$P_s$ would reduce due to decays of atoms from Rydberg states.

As the approximation made for the effective Hamiltonian also causes
some errors in the operation, we here investigate the robustness
against systematic errors of the implementation of the Not gate with
the full Hamiltonian. The final average fidelities $\bar{F}_N(T)$
versus $\epsilon$ with $\chi_0=1$, $\chi_0=0.5$ and $\chi_0=0$ are
plotted in Fig.~\ref{fig4}(b) based on the evolution governed by the
full Hamiltonian. We can see from the red-dotted line in
Fig.~\ref{fig4}(b) that, when $\epsilon\in[-0.1,0.1]$ (corresponding
to $\tilde{\epsilon}\in[-0.2,0.2]$), the average fidelity with
$\chi_0=1$ keeps higher than 0.988. Therefore, the robustness
against systematic errors is inherited when the full dynamics being
considered. Moreover, seen from the green-solid line and the
blue-dashed line in Fig.~\ref{fig4}(b), the average fidelities with
$\chi_0=0$ and $\chi_0=0.5$ is much more sensitive to systematic
errors. For example, for $\chi_0=0$, the average fidelity is only
$\bar{F}_N(T)=0.8089$ when $\epsilon=0.1$.

In the end of this section, let us study the influence of
dissipation with the full Hamiltonian. As the evolution is not
unitary when dissipation taken into account, we consider the
evolution with the initial state $|r0\rangle_{01}$ as example to
show evolution governed by the master equation. We plot the fidelity
of the evolution before measuring the state of the auxiliary atom as
$F_N(t)=\mathrm{Tr}[U_N\rho(0)U^\dag_N\rho(t)]$ versus $t$ in
Fig.~\ref{fig4}(c). As shown by the red-solid line in
Fig.~\ref{fig4}(c), the final fidelity without the decay is
$F_N(T)|_{\gamma=0}=0.9933$. In addition, the green-dotted line
points out the final fidelity calculated by Eq.~(\ref{de2}) with
$\gamma=1$kHz is
$\tilde{F}_N(T)|_{\gamma=1\mathrm{kHz}}=\exp(-\gamma
T)|_{\gamma=1\mathrm{kHz}}\times F_N(T)|_{\gamma=0}=0.9708$.
Moreover, the blue-dashed line indicates the final fidelity given by
the full master equation (replacing the effective Hamiltonian
$H_e(t)$ in Eq.~(\ref{de1}) by the full Hamiltonian $H(t)$) with
$\gamma=1$kHz is $F_N(T)|_{\gamma=1\mathrm{kHz}}=0.9704$, which is
in accordance with the theoretical result
$\tilde{F}_N(T)|_{\gamma=1\mathrm{kHz}}$. Moreover, by comparing the
green-dotted line and the blue-dashed line, we can also find the
dynamics governed by the full master equation matches well with that
discussed in Sec. IIIB in the twelve-dimensional subspace. The
result also proves the theoretical analysis of the effective
Hamiltonian and the subspace is valid. In Fig.~\ref{fig4}(d), we
consider the fidelity with broader range of decay rate $\gamma$,
where the fidelity obtained from full master equation is a little
higher than that estimated from Eq.~(\ref{de2}) when dissipation is
relatively strong. This is because dissipation also restrains the
population of $|rr\rangle_{01}$ alongside with the Rydberg blockade.
When $\gamma$ reaches 4kHz, the obtained fidelity is 0.9079, still
high than 0.9. Therefore, the protocol holds robustness against
dissipation. We also examine the successful probability for
measuring the Rydberg state of the auxiliary atom as
$P_s=\mathrm{Tr}[\mathcal{P}_r\rho(T)]$ with the projection operator
$\mathcal{P}_r=|r\rangle_0\langle r|\otimes\mathbbold{1}_{1}$
($\mathbbold{1}_{1}$ is the identity operator for computational atom
1), and the result is shown in Table I. Furthermore, the fidelity
after successful measurement of the auxiliary atom and the purity of
the density operator of computational atom 1 as
$F_N'=\mathrm{Tr}[\mathcal{P}_r\rho(T)\mathcal{P}_rU_N\rho(0)U^\dag_N]/P_s$
and
$\varrho_1=\mathrm{Tr}[\mathcal{P}_r\rho(T)\mathcal{P}_r\rho(T)]/P_s^2$
are also investigated, respectively. According to the data in Table
I, although the successful probability decreases when the decay rete
increases, after successful measurement of the auxiliary atom, the
fidelity of the Not gate and the purity of the density operator of
atom 1 are changed very slightly. Even when $\gamma=4$kHz, the
fidelity $F_N'$ is only reduced about 0.0004, and the purity
$\varrho_1$ is still 0.9991. Therefore, by using the protocol, we
can still obtain nearly perfect unitary evolution in the presence of
dissipation if the measurement result of the state of the auxiliary
atom is $|r\rangle_0$, which accords with the theoretical analysis
in Sec. IIIB.
\begin{center}
{\small\begin{tabular}{c|ccccc} \multicolumn{6}{c}{Table I.
Successful probability $P_s$, fidelity $F_N'$}\\
\multicolumn{6}{c}{and purity $\varrho_1$ with different decay
rates.}\\
\hline\hline \ \ \ \ $\gamma$ (kHz)\ \ \ &\ \ \ 0\ \ \ &\ \ \ 1\ \ \ &\ \ \ 2\ \ \ &\ \ \ 3\ \ \ &\ \ \ 4\ \ \ \\
\hline \ \ \ $P_s$\ \ \ &\ \ \ 0.9988\ \ \ &\ \ \ 0.9770\ \ \ &\ \ \ 0.9557\ \ \ &\ \ \ 0.9348\ \ \ &\ \ \ 0.9144\ \ \ \\
\hline \ \ \ $F_N'$\ \ \ &\ \ \ 0.9933\ \ \ &\ \ \ 0.9932\ \ \ &\ \ \ 0.9931\ \ \ &\ \ \ 0.9930\ \ \ &\ \ \ 0.9929\ \ \ \\
\hline \ \ \ $\varrho_1$\ \ \ &\ \ \ 1.0000\ \ \ &\ \ \ 0.9998\ \ \ &\ \ \ 0.9996\ \ \ &\ \ \ 0.9993\ \ \ &\ \ \ 0.9991\ \ \ \\
\hline\hline
\end{tabular}}
\end{center}

\subsection{Numerical analysis of two-qubit entangling gate}

\begin{figure}
\scalebox{0.6}{\includegraphics[scale=0.7]{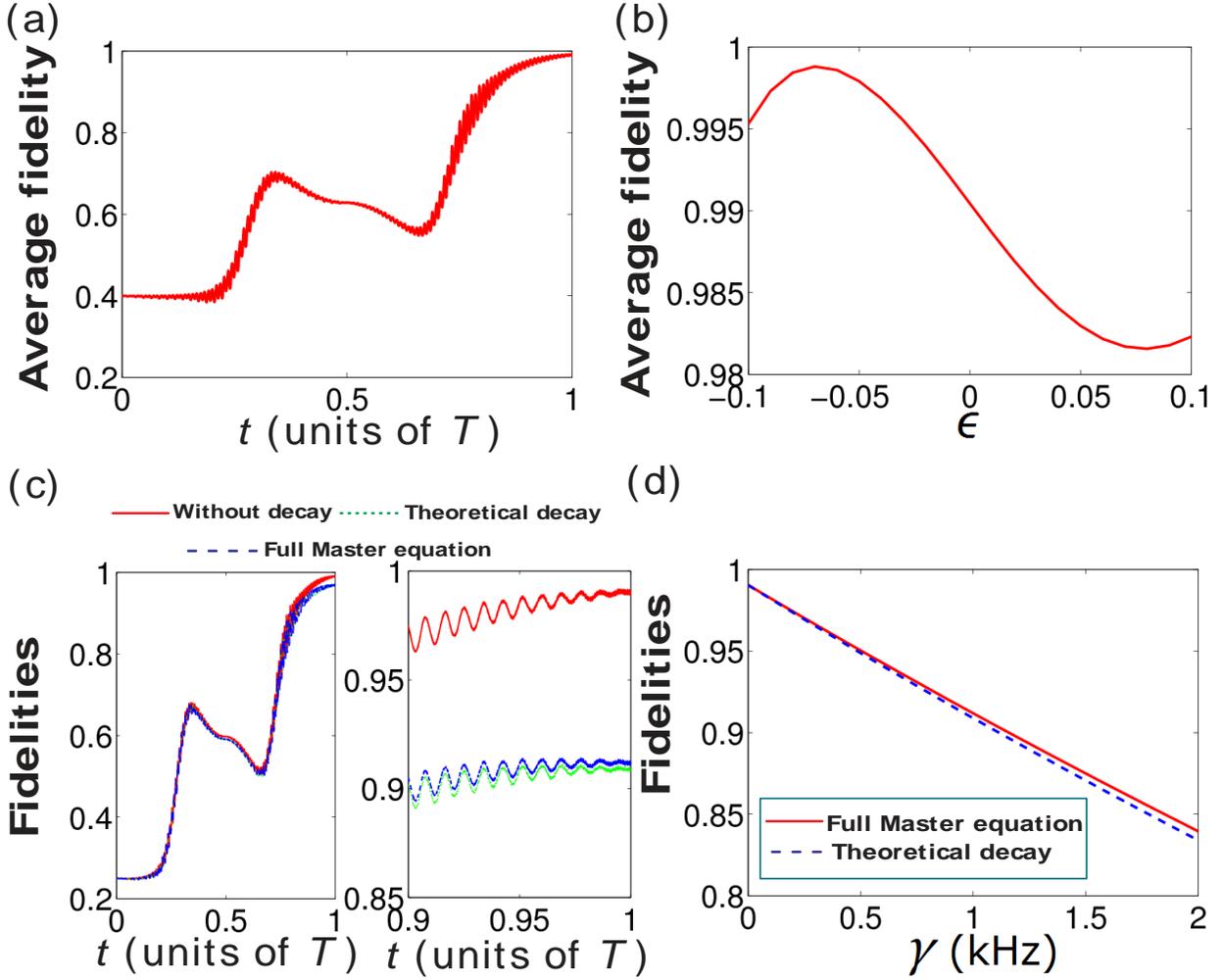}} \caption{(a)
Average fidelity $\bar{F}_{CN}(t)$ of the implementation of the
C-Not gate versus $t$ with Full Hamiltonian. (b) The final average
fidelities $\bar{F}_{CN}(T)$ versus $\epsilon$. (c) Fidelities
$F_{CN}(t)$ of the implementation of the C-Not gate versus $t$
calculated by the master equation with initial state
$(|r00\rangle_{012}+|r10\rangle_{012})/\sqrt{2}$. Red-solid line:
$\gamma=0$, green-dotted line: $\gamma=1$kHz with theoretical result
in Eq.~(\ref{de2}), blue-dashed line: $\gamma=1$kHz with the full
master equation. (d) Final fidelities $F_{CN}(T)$ of the
implementation of the C-Not gate versus decay rate $\gamma$ with
initial state $(|r00\rangle_{012}+|r10\rangle_{012})/\sqrt{2}$ by
using theoretical result in Eq.~(\ref{de2}) (blue-dashed line) and
the full master equation (red-solid line).}\label{fig5}
\end{figure}

We now make numerical analysis of two-qubit entangling gate. As an
example to show the implementation of two-qubit entangling gate, we
amply analyze the realization of the C-Not gate in the following
discussions. In this case, we switch off laser pulses
$\Omega_{1}(t)$, $\Omega_{1}'(t)$, $\Omega_{01}(t)$,
$\Omega_{01}'(t)$, $\Omega_{23}(t)$ and $\Omega_{23}'(t)$ for
$\Omega_{e_1}(t)=0$ and $\theta_1=0$. Besides, the wave form of
$\Omega_{e2}(t)$ is considered the same as that discussed in Sec. IV
with invariant-based reverse engineering and the
systematic-error-nullification method. To meet the condition
$\tilde{\Omega}_3\gg|\Omega_{e2}(t)|$, we consider
$\tilde{\Omega}_3=100/T$. Moreover, the Rydberg interaction strength
and detunings are set as $V=27000/T$, $\Delta_2=360/T$, and
$\Delta_3=1500/T$ to build up the effective Hamiltonian. In this
case, the total operation time is $85.94\mu$s with
$V=2\pi\times50$MHz, and the successful probability is $P_s=91.76\%$
with $\gamma=1$kHz. In addition, we eliminate the Stark shifts as
\begin{eqnarray}\label{e27}
-\frac{2V}{V^2-\Delta_2^2}&\times&[\bar{\Omega}_2^2(|r\pm+\rangle_{012}\langle
r\pm+|+|0r+\rangle_{012}\langle
0r+|)\cr\cr&&+\bar{\Omega}_{02}^2(|0r\pm\rangle_{012}\langle0r\pm|+|0\pm
r\rangle_{012}\langle0\pm r|)]\cr\cr
-\frac{2V}{V^2-\Delta_3^2}&\times&[\bar{\Omega}_{13}^2(|r+\pm\rangle_{012}\langle
r+\pm|+|0+r\rangle_{012}\langle
0+r|)\cr\cr&&+\bar{\Omega}_{23}^2(|r\pm+\rangle_{012}\langle
r\pm+|+|0r+\rangle_{012}\langle 0r+|)],
\end{eqnarray}
by using auxiliary pulses \cite{SSLPRA96,LDXPRA98}. The average
fidelity of the implementation of the C-Not gate is defined as
\cite{ZanardiPRA70,PedersenPLA367}
\begin{eqnarray}\label{e28}
\bar{F}_{CN}(t)=\frac{1}{\mathcal{N}'(\mathcal{N}'+1)}\{\mathrm{Tr}[M'(t)M^{\prime\dag}(t)]
+|\mathrm{Tr}[M'(t)]|^2\},
\end{eqnarray}
with $M'(t)=|r\rangle_0\langle r|\otimes U_{CN}+|0\rangle_0\langle
0|\otimes\mathbbold{1}_{12}$ and $\mathcal{N}'=4$ being the
dimension of the computational subspace. We plot $\bar{F}_{CN}(t)$
versus $t$ in Fig.~\ref{fig5}(a) and obtain the average fidelity of
the implementation of the C-Not gate as $\bar{F}_{CN}(T)=0.9904$ at
$t=T$. Therefore, the C-Not gate can be successfully realized with
the protocol.

We also check the robustness of the implementation of the C-Not gate
against systematic errors of laser pulses. The final average
fidelity $\bar{F}_{CN}(T)$ of the implementation of the C-Not gate
versus error coefficient $\epsilon$ is plotted in
Fig.~\ref{fig5}(b). Seen from Fig.~\ref{fig5}(b), $\bar{F}_{CN}(T)$
is always higher than 0.9823 when $\epsilon\in[-0.1,0.1]$.
Therefore, the implementation of the C-Not gate is also insensitive
to systematic errors. Moreover, we can also see from
Fig.~\ref{fig5}(b) that errors with $\epsilon>0$ may decrease the
average fidelity $\bar{F}_{CN}(T)$ in a range. However, the errors
with $\epsilon<0$ may increase the average fidelity
$\bar{F}_{CN}(T)$ in a range on the contrary. This is because the
satisfaction of the condition
$\tilde{\Omega}_3\gg\tilde{\Omega}_{e_2}(t)$ becomes worse when
$\epsilon>0$. Although the deviations of $\Omega_{e_2}(t)$ and
$\Omega_{e_3}(t)$ are both about $2\epsilon$ to the origin ones
according to Eq.~(\ref{e7}), according to the second-order
perturbation theory, the coefficients of error terms caused by the
terms with high-frequency oscillations is approximately increased
from $\Omega_{e_2}^2(t)/\tilde{\Omega}_3$ to
$(1+2\epsilon)^2\Omega_{e_2}^2(t)/[(1+2\epsilon)\tilde{\Omega}_3]=(1+2\epsilon)\Omega_{e_2}^2(t)/\tilde{\Omega}_3$.
Therefore, when $\epsilon<0$, part of errors caused by the
approximation $\tilde{\Omega}_3\gg\tilde{\Omega}_{e_2}(t)$ are
compensated by systematic errors. The maxima of
$\bar{F}_{CN}(T)=0.9988$ appears at $\epsilon=-0.07$, which may be
considered as a correction in the design of pulses for higher
fidelity.

Finally, we check the performance of the protocol under the
influence of the dissipation by considering the initial state
$(|r00\rangle_{012}+|r10\rangle_{012})/\sqrt{2}$ in the evolution
governed by the master equation. We plot the fidelity of the
evolution before measuring the state of the auxiliary atom as
$F_{CN}(t)=\mathrm{Tr}[U_{CN}\rho(0)U^\dag_{CN}\rho(t)]$ versus $t$
in Fig.~\ref{fig5}(c). Seen from the red-solid line in
Fig.~\ref{fig5}(c), the fidelity of the implementation of the C-Not
gate is $F_{CN}(T)|_{\gamma=0}=0.9904$ at $t=T$. Moreover, the
green-dotted line in Fig.~\ref{fig4}(c) is the fidelity calculated
by Eq.~(\ref{de2}) with $\gamma=1$kHz, which gives
$\tilde{F}_{CN}(T)|_{\gamma=1\mathrm{kHz}}=\exp(-\gamma
T)|_{\gamma=1\mathrm{kHz}}\times F_N(T)|_{\gamma=0}=0.9089$. In
addition, the blue-dashed line indicates that the final fidelity
obtained by the full master equation is
$F_{CN}(T)|_{\gamma=1\mathrm{kHz}}=0.9115$. We also plot the final
fidelity $F_{CN}(T)$ of the implementation of the C-Not gate with
the initial state $(|r00\rangle_{012}+|r10\rangle_{012})/\sqrt{2}$
versus decay rate $\gamma$ in Fig.~\ref{fig5}(d), where one can also
find that the final fidelity $F_{CN}(T)$ plotted by the full master
equation is higher than the theoretical result obtained by
Eq.~(\ref{de2}) when the dissipation is relatively strong. This is
because populations of states with multiple atoms in Rydberg states
are restrained by dissipation. In the implementation of the
two-qubit gate, the fidelity decreases more significantly compared
with the result in the implementation of the single-qubit gate when
dissipation is taken into account due to the increase of the total
operation time. However, by measuring the state of the auxiliary
atom, influence of dissipation can be removed if the result is
$|r\rangle_0$. Here, we define the successful probability of getting
the result $|r\rangle_0$ as
$P_s'=\mathrm{Tr}[\mathcal{P}_r'\rho(T)]$ with
$\mathcal{P}_r'=|r\rangle_0\langle r|\times\mathbbold{1}_{12}$.
Furthermore, the fidelity of the C-Not gate and the purity of the
density operator of the computational atoms 1 and 2 after successful
measurement of auxiliary atom are defined as
$F_{CN}'=\mathrm{Tr}[\mathcal{P}_r'\rho(T)\mathcal{P}_r'U_{CN}\rho(0)U^\dag_{CN}]/P_s$
and
$\varrho_{12}=\mathrm{Tr}[\mathcal{P}_r'\rho(T)\mathcal{P}_r'\rho(T)]/P_s^2$,
respectively. We calculate $P_s'$, $F_{CN}'$ and $\varrho_{12}$ with
some samples shown in Table II. According to the data in Table II,
although the successful probability $P_s'$ decreases with the
increase of the decay rate $\gamma$, the fidelity $F_{CN}'$ and the
purity $\varrho_{12}$ in the case of successful measurement are
almost unchanged. For example, when $\gamma=1$kHz, the decreases of
$F_{CN}'$ and $\varrho_{12}$ are only 0.0005 and 0.0012,
respectively. In addition, the measurement of state $|r\rangle_0$
can also help to reduce the effect of systematic errors and errors
caused by the approximation. Because these errors would also make
the final state of auxiliary atom deviate from $|r\rangle_0$. For
example, when $\gamma=0$, compared with the fidelity
$F_{CN}(T)|_{\gamma=0}=0.9904$ obtained without the measurement of
auxiliary atom, the fidelity after successful measurement is
improved to $F_{CN}'=0.9992$. From the results above, the protocol
is helpful to maintain a nearly perfect two-qubit unitary operation
in the dissipative environment.
\begin{center}
{\small\begin{tabular}{c|ccc} \multicolumn{4}{c}{Table II.
Successful probability $P_s'$, fidelity $F_{CN}'$}\\
\multicolumn{4}{c}{and the purity $\varrho_{12}$ with different
decay rates.}\\
\hline\hline \ \ \ \ $\gamma$ (kHz)\ \ \ &\ \ \ 0\ \ \ &\ \ \ 1\ \ \ &\ \ \ 2\ \ \ \\
\hline \ \ \ $P_s'$\ \ \ &\ \ \ 0.9911\ \ \ &\ \ \ 0.9130\ \ \ &\ \ \ 0.8411\ \ \ \\
\hline \ \ \ $F_{CN}'$\ \ \ &\ \ \ 0.9992\ \ \ &\ \ \ 0.9987\ \ \ &\ \ \ 0.9981\ \ \ \\
\hline \ \ \ $\varrho_{12}$\ \ \ &\ \ \ 0.9995\ \ \ &\ \ \ 0.9983\ \ \ &\ \ \ 0.9971\ \ \ \\
\hline\hline
\end{tabular}}
\end{center}

\section{Conclusion}

In conclusion, we have proposed a protocol to realize atomic
nonadiabatic holonomic quantum computation (NHQC) in the regime of
Rydberg blockade. Assisted by the strong interaction between Rydberg
atoms, the effective Hamiltonian was built up by the second-order
perturbation theory with proper detunings. Based on the derived
effective Hamiltonian, we further designed the laser pulses with the
help of invariant-based reverse engineering. The advantages of
invariant-based reverse engineering in NHQC have been shown in the
protocol. On one hand, eigenvectors of the dynamic invariant provide
natural evolution paths for NHQC. On the other hand, invariant-based
reverse engineering is also compatible with the
systematic-error-sensitivity nullified method, which makes the
evolution insensitive to systematic errors of laser pulses.
Generally, there are many different choices for the parameters in
the nullification of systematic-error-sensitivity $Q_s$. In the
implementation of a single- or two-qubit gate, by nullifying the
systematic-error-sensitivity $Q_s$, we can derive optimal solutions
for Rabi frequencies of laser pulses that can make the
implementation insensitive to the systematic errors of pulses.
Therefore, in the implementation of a general quantum circuit
composed of a sequence of single- and two-qubit gates, we need to
individually derive the Rabi frequencies of pulses with $Q_s$ being
nullified for each gate in the sequence. As a result, we can obtain
a sequence of Rabi frequencies of pulses for each step of
operations. In this way, we make the total
systematic-error-sensitivity $Q_s$ nullified in the whole process,
so that the implementation of the quantum circuit maintains a high
level when there exist systematic errors. Moreover, in the design of
pulses, by setting proper boundary conditions for the time
derivatives of control parameters, we can make each pulse in the
sequence vanishes at the final time of each step. Pulses in each two
adjacent steps can be connected as composite pulse described by a
continuous function. Therefore, the composite pulse does not involve
sudden changes in the whole process. In a real implementation, we
just need to apply the composite pulse to the system. In addition,
we analyzed the evolution in a dissipative environment based on the
master equation. Both the theoretical and numerical results showed
that the protocol can realize nearly perfect unitary operations if
the auxiliary atom in the Rydberg state is successfully measured.
Considering a typical decay rate $\gamma=1$kHz of a Rydberg state,
the protocol produces acceptable successful probabilities of
measurements as 0.9770 and 0.9130 for single- and two-qubit gates,
respectively. Compared with previous Rydberg-atom-based NHQC
protocol \cite{KYHPRA972}, the protocol has several advantages.
Firstly, with the invariant-based reverse engineering in the
protocol, we obtain an invariant of the system, whose eigenvectors
can be used as paths for NHQC by only eliminating the acquired
dynamic phases. This makes the parameter selections more convenient
compared with that in protocol \cite{KYHPRA972}, where both the
parallel transport conditions and the unavailable couplings should
be considered. Secondly, as a result of the convenience of parameter
selections, the protocol can incorporate with the
systematic-error-sensitivity nullified method. This makes the
protocol maintain high fidelities when systematic errors appear.
Thirdly, the system in protocol \cite{KYHPRA972} will be in a mixed
state when dissipation exists, while in the current protocol, as the
heralded implementation is considered, the state of the system
nearly maintains in pure state with successful measurement on the
auxiliary atom. As the protocol is fully compatible with the
advantages of geometric phases, reverse engineering,
systematic-error-sensitivity nullified method, heralded
implementation, and Rydberg interaction, we hope the protocol can be
helpful for the precise quantum computation in dissipative
environment.

\section*{Acknowledgement}

This work was supported by the National Natural Science Foundation
of China under Grants No. 11805036.

\section*{Appendix A: Derivations of the effective Hamiltonian}

For the atomic system shown in Sec. IIIA, the Hamiltonian of the
whole system under the rotating-wave approximation reads
\begin{eqnarray}\label{fp1}
&&H(t)=H_1(t)+H_2(t)+H_3(t)+H_\mathrm{v},\cr\cr
&&H_1(t)=\sum\limits_{j,k=1}^2[\Omega_{kj}(t)e^{i\Delta_{k}t}+\Omega_{kj}'(t)e^{-i\Delta_{k}t}]|j-1\rangle_k\langle
r|+\mathrm{H.c.},\cr\cr
&&H_2(t)=\sum\limits_{k=1}^2[\Omega_{0k}(t)e^{i\Delta_{k}t}+\Omega_{0k}'(t)e^{-i\Delta_{k}t}]|0\rangle_0\langle
r|+\mathrm{H.c.},\cr\cr
&&H_3(t)=\sum\limits_{j,k=1}^2[\Omega_{kj3}(t)e^{i\Delta_{3}t}+\Omega_{kj3}'(t)e^{-i\Delta_{3}t}]|j-1\rangle_k\langle
r|+\mathrm{H.c.},\cr\cr &&H_{\mathrm{v}}=V(|rr\rangle_{01}\langle
rr|+|rr\rangle_{02}\langle rr|+|rr\rangle_{12}\langle rr|).
\end{eqnarray}
With the assumptions in Eq.~(\ref{fp00}), the Hamiltonian in
Eq.~(\ref{fp1}) can be rewritten by
\begin{eqnarray}\label{fp2}
&&H(t)=H_1(t)+H_2(t)+H_3(t)+H_\mathrm{v},\cr\cr
&&H_1(t)=\sum\limits_{k=1}^2[\Omega_{k}(t)e^{i\Delta_{k}t}+\Omega_{k}'(t)e^{-i\Delta_{k}t}]|+\rangle_k\langle
r|+\mathrm{H.c.},\cr\cr
&&H_2(t)=\sum\limits_{k=1}^2[\Omega_{0k}(t)e^{i\Delta_{k}t}+\Omega_{0k}'(t)e^{-i\Delta_{k}t}]|0\rangle_0\langle
r|+\mathrm{H.c.},\cr\cr
&&H_3(t)=\sum\limits_{k=1}^2[\Omega_{k3}(t)e^{i\Delta_{3}t}+\Omega_{k3}'(t)e^{-i\Delta_{3}t}]|+\rangle_k\langle
r|+\mathrm{H.c.},\cr\cr &&H_{\mathrm{v}}=V(|rr\rangle_{01}\langle
rr|+|rr\rangle_{02}\langle rr|+|rr\rangle_{12}\langle rr|).
\end{eqnarray}
In Eq.~(\ref{fp2}), $|+\rangle_k$ reads
$|+\rangle_k=\cos(\theta_k/2)|0\rangle_k+\sin(\theta_k/2)e^{i\varphi_k}|1\rangle_k$,
and it has an orthogonal partner as
$|-\rangle_k=\sin(\theta_k/2)|0\rangle_k-\cos(\theta_k/2)e^{i\varphi_k}|1\rangle_k$.
We assume that the system works at the Rydberg blockade regime with
$V\gg\{\Delta_k,\Delta_3,|\Omega_k(t)|,|\Omega_k'(t)|,|\Omega_{0k}(t)|,|\Omega_{0k}'(t)|,|\Omega_{k3}(t)|,|\Omega_{k3}'(t)|\}$,
the Hamiltonian in the rotating frame of
$U_0(t)=\exp(-iH_\mathrm{v}t)$ can be derived as
\begin{eqnarray}\label{fp3}
\tilde{H}(t)&=&\tilde{H}_1(t)+\tilde{H}_2(t)+\tilde{H}_3(t),\cr\cr
\tilde{H}_1(t)&=&[\Omega_{1}(t)e^{i\Delta_1t}+\Omega_{1}'(t)e^{-i\Delta_1t}](|0++\rangle_{012}\langle
0r+|+|0+-\rangle_{012}\langle0r-|)\cr\cr
&+&[\Omega_{2}(t)e^{i\Delta_2t}+\Omega_{2}'(t)e^{-i\Delta_2t}](|0++\rangle_{012}\langle
0+r|+|0-+\rangle_{012}\langle0-r|)+\mathrm{H.c.},\cr\cr
\tilde{H}_2(t)&=&\sum\limits_{k=1}^2\sum\limits_{\jmath,\jmath'=\pm}[\Omega_{0k}(t)e^{i\Delta_kt}
+\Omega_{0k}'(t)e^{-i\Delta_kt}]|0\jmath\jmath'\rangle_{012}\langle
r\jmath\jmath'|+\mathrm{H.c.},\cr\cr
\tilde{H}_3(t)&=&[\Omega_{13}(t)e^{i\Delta_3t}+\Omega_{13}'(t)e^{-i\Delta_3t}](|0++\rangle_{012}\langle
0r+|+|0+-\rangle_{012}\langle0r-|)\cr\cr
&+&[\Omega_{23}(t)e^{i\Delta_3t}+\Omega_{23}'(t)e^{-i\Delta_3t}](|0++\rangle_{012}\langle
0+r|+|0-+\rangle_{012}\langle0-r|)+\mathrm{H.c.},
\end{eqnarray}
by omitting the terms with oscillation frequencies in the scale of
$V$. To further simplify the dynamics of the system, we consider the
condition
$\{\Delta_k,\Delta_3,|\Delta_1-\Delta_2|,|\Delta_1-\Delta_3|,|\Delta_2-\Delta_3|\}\gg\{|\Omega_k(t)|,|\Omega_k'(t)|,|\Omega_{0k}(t)|,|\Omega_{0k}'(t)|,|\Omega_{k3}(t)|,|\Omega_{k3}'(t)|\}$.
With the help of the second-order perturbation theory
\cite{JamesCJP85}, the effective Hamiltonian is derived as
\begin{eqnarray}\label{fp4}
H_e(t)&=&H_{e_0}(t)+H_{e_1}(t)+H_{e_2}(t)+H_{e_3}(t),\cr\cr
H_{e_0}(t)&=&\Omega_{e_1}(t)|r++\rangle_{012}\langle0r+|
+\Omega_{e_2}(t)|r++\rangle_{012}\langle0+r|+\mathrm{H.c.}\cr\cr
H_{e_1}(t)&=&\Omega_{e_1}(t)|r+-\rangle_{012}\langle
0r-|+\mathrm{H.c.},\cr\cr
H_{e_2}(t)&=&\Omega_{e_2}(t)|r-+\rangle_{012}\langle
0-r|+\mathrm{H.c.},\cr\cr
H_{e_3}(t)&=&\Omega_{e_3}(t)|0r+\rangle_{012}\langle0+r|+\mathrm{H.c.}
\end{eqnarray}
with
\begin{eqnarray}\label{fp5}
\Omega_{e_k}(t)=\frac{\Omega_k'(t)\Omega_{0k}^{\prime*}(t)-\Omega_k(t)\Omega_{0k}^*(t)}{\Delta_k},\
\Omega_{e_3}(t)=\frac{\Omega_{13}'(t)\Omega_{23}^{\prime*}(t)-\Omega_{13}(t)\Omega_{23}^*(t)}{\Delta_3}.
\end{eqnarray}
With the assumptions in Eq.~(\ref{fp01}), $\Omega_{e_k}(t)$ can be
simplified as
\begin{eqnarray}\label{fp6}
&&\Omega_{e_k}(t)=\frac{2\bar{\Omega}_k(t)\bar{\Omega}_{0k}(t)}{\Delta_k}\sin(\frac{\mu_k-\mu_k'}{2})e^{i(\mu_k'+\mu_k+\pi)/2},\cr\cr
&&\Omega_{e_3}(t)=\frac{2\bar{\Omega}_{13}(t)\bar{\Omega}_{23}(t)}{\Delta_k}\sin(\frac{\mu_3-\mu_3'}{2})e^{i(\mu_3'+\mu_3+\pi)/2}.
\end{eqnarray}
Considering $\mu_k-\mu_k'=\pi$ and $\mu_3-\mu_3'=\pi$, we have
$\Omega_{e_k}(t)=\tilde{\Omega}_{k}(t)e^{i\mu_k(t)}$,
$\Omega_{e_3}(t)=\tilde{\Omega}_{3}(t)e^{i\mu_3(t)}$ with
\begin{eqnarray}\label{fp7}
\tilde{\Omega}_{k}(t)=\frac{2\bar{\Omega}_k(t)\bar{\Omega}_{0k}(t)}{\Delta_k},\
\tilde{\Omega}_{3}(t)=\frac{2\bar{\Omega}_{13}(t)\bar{\Omega}_{23}(t)}{\Delta_3}.
\end{eqnarray}

\section*{Appendix B: Matrix elements of time derivative of the density operator}

For simplicity, we number the basis vectors of subspace
$\mathcal{B}'$ in Table III.

\begin{center}
{\small\begin{tabular}{cccc} \multicolumn{4}{c}{Table III. Basis vector of subspace $\mathcal{B}'$.}\\
\hline\hline
\ \ \ \ $|\Phi_1\rangle$\ \ \ \ &\ \ \ \ $|\Phi_2\rangle$\ \ \ \ & \ \ \ \ $|\Phi_3\rangle$\ \ \ \ &\ \ \ \ $|\Phi_4\rangle$\ \ \ \ \\
\hline
\ \ \ \ $|r++\rangle_{012}$\ \ \ \ &\ \ \ \ $|r+-\rangle_{012}$\ \ \ \ & \ \ \ \ $|r-+\rangle_{012}$\ \ \ \ &\ \ \ \ $|r--\rangle_{012}$\ \ \ \ \\
\hline\hline
\ \ \ \ $|\Phi_5\rangle$\ \ \ \ &\ \ \ \ $|\Phi_6\rangle$\ \ \ \ & \ \ \ \ $|\Phi_7\rangle$\ \ \ \ &\ \ \ \ $|\Phi_8\rangle$\ \ \ \ \\
\hline
\ \ \ \ $|0r+\rangle_{012}$\ \ \ \ &\ \ \ \ $|0r-\rangle_{012}$\ \ \ \ & \ \ \ \ $|0+r\rangle_{012}$\ \ \ \ &\ \ \ \ $|0-r\rangle_{012}$\ \ \ \ \\
\hline\hline
\ \ \ \ $|\Phi_9\rangle$\ \ \ \ &\ \ \ \ $|\Phi_{10}\rangle$\ \ \ \ & \ \ \ \ $|\Phi_{11}\rangle$\ \ \ \ &\ \ \ \ $|\Phi_{12}\rangle$\ \ \ \ \\
\hline
\ \ \ \ $|0++\rangle_{012}$\ \ \ \ &\ \ \ \ $|0+-\rangle_{012}$\ \ \ \ & \ \ \ \ $|0-+\rangle_{012}$\ \ \ \ &\ \ \ \ $|0--\rangle_{012}$\ \ \ \ \\
\hline \hline
\end{tabular}}
\end{center}

According to Eq.~(\ref{de1}), the time derivatives of nonzero matrix
elements of $\rho(t)$ can be calculated as
\begin{eqnarray}\label{apa1}
&&\dot{\rho}_{1,1}=-i\Omega_{e_1}\rho_{5,1}+i\Omega_{e_1}^*\rho_{1,5}-i\Omega_{e_2}\rho_{7,1}+i\Omega_{e_2}^*\rho_{1,7}-\gamma\rho_{1,1},\cr\cr
&&\dot{\rho}_{1,2}=-\gamma\rho_{1,2},\ \
\dot{\rho}_{1,3}=-\gamma\rho_{1,3}\ \
\dot{\rho}_{1,4}=-\gamma\rho_{1,4},\cr\cr
&&\dot{\rho}_{1,5}=i\Omega_{e_1}(\rho_{1,1}-\rho_{5,5})-\gamma\rho_{1,5},\
\ \dot{\rho}_{1,6}=-\gamma\rho_{1,6},\cr\cr
&&\dot{\rho}_{1,7}=i\Omega_{e_2}(\rho_{1,1}-\rho_{7,7})-\gamma\rho_{1,7},\
\ \dot{\rho}_{1,8}=-\gamma\rho_{1,8},\cr\cr
&&\dot{\rho}_{2,1}=-\gamma\rho_{2,1},\ \
\dot{\rho}_{2,2}=-i\Omega_{e_1}\rho_{6,2}+i\Omega_{e_1}^*\rho_{2,6}-\gamma\rho_{2,2},\cr\cr
&&\dot{\rho}_{2,3}=-\gamma\rho_{2,3},\ \
\dot{\rho}_{2,4}=-\gamma\rho_{2,4},\ \
\dot{\rho}_{2,5}=-\gamma\rho_{2,5},\cr\cr
&&\dot{\rho}_{2,6}=i\Omega_{e_1}(\rho_{2,2}-\rho_{6,6})-\gamma\rho_{2,6},\
\ \dot{\rho}_{2,7}=-\gamma\rho_{2,7},\ \
\dot{\rho}_{2,8}=-\gamma\rho_{2,8},\cr\cr
&&\dot{\rho}_{3,1}=-\gamma\rho_{3,1},\ \
\dot{\rho}_{3,2}=-\gamma\rho_{3,2},\ \
\dot{\rho}_{3,3}=-i\Omega_{e_2}\rho_{8,3}+i\Omega_{e_2}^*\rho_{3,8}-\gamma\rho_{3,3}\cr\cr
&&\dot{\rho}_{3,4}=-\gamma\rho_{3,4},\ \
\dot{\rho}_{3,5}=-\gamma\rho_{3,5},\ \
\dot{\rho}_{3,6}=-\gamma\rho_{3,6},\cr\cr
&&\dot{\rho}_{3,7}=-\gamma\rho_{3,7},\ \
\dot{\rho}_{3,8}=i\Omega_{e_2}(\rho_{3,3}-\rho_{8,8})-\gamma\rho_{3,8},\cr\cr
&&\dot{\rho}_{4,1}=-\gamma\rho_{4,1},\ \
\dot{\rho}_{4,2}=-\gamma\rho_{4,2},\ \
\dot{\rho}_{4,3}=-\gamma\rho_{4,3},\ \
\dot{\rho}_{4,4}=-\gamma\rho_{4,4},\cr\cr
&&\dot{\rho}_{4,5}=-\gamma\rho_{4,5},\ \
\dot{\rho}_{4,6}=-\gamma\rho_{4,6},\ \
\dot{\rho}_{4,7}=-\gamma\rho_{4,7},\ \
\dot{\rho}_{4,8}=-\gamma\rho_{4,8},\cr\cr
&&\dot{\rho}_{5,1}=-i\Omega_{e_1}^*(\rho_{1,1}-\rho_{5,5})-\gamma\rho_{5,1},\
\ \dot{\rho}_{5,2}=-\gamma\rho_{5,2},\ \
\dot{\rho}_{5,3}=-\gamma\rho_{5,3},\cr\cr
&&\dot{\rho}_{5,4}=-\gamma\rho_{5,4},\ \
\dot{\rho}_{5,5}=i\Omega_{e_1}\rho_{5,1}-i\Omega_{e_1}^*\rho_{1,5}-i\Omega_{e_3}\rho_{7,5}+i\Omega_{e_3}^*\rho_{5,7}-\gamma\rho_{5,5},\cr\cr
&&\dot{\rho}_{5,6}=-\gamma\rho_{5,6},\ \
\dot{\rho}_{5,7}=i\Omega_{e_3}(\rho_{5,5}-\rho_{7,7})-\gamma\rho_{5,7},\
\ \dot{\rho}_{5,8}=-\gamma\rho_{5,8},\cr\cr
&&\dot{\rho}_{6,1}=-\gamma\rho_{6,1},\ \
\dot{\rho}_{6,2}=-i\Omega_{e_1}^*(\rho_{2,2}-\rho_{6,6})-\gamma\rho_{6,2},\cr\cr
&&\dot{\rho}_{6,3}=-\gamma\rho_{6,3},\ \
\dot{\rho}_{6,4}=-\gamma\rho_{6,4},\ \
\dot{\rho}_{6,5}=-\gamma\rho_{6,5},\cr\cr
&&\dot{\rho}_{6,6}=i\Omega_{e_1}\rho_{6,2}-i\Omega_{e_1}^*\rho_{2,6}-\gamma\rho_{6,6},\
\ \dot{\rho}_{6,7}=-\gamma\rho_{6,7},\ \
\dot{\rho}_{6,8}=-\gamma\rho_{6,8},\cr\cr
&&\dot{\rho}_{7,1}=-i\Omega_{e_2}^*(\rho_{1,1}-\rho_{7,7})-\gamma\rho_{7,1},\
\ \dot{\rho}_{7,2}=-\gamma\rho_{7,2},\ \
\dot{\rho}_{7,3}=-\gamma\rho_{7,3},\cr\cr
&&\dot{\rho}_{7,4}=-\gamma\rho_{7,4},\ \
\dot{\rho}_{7,5}=-i\Omega_{e_3}^*(\rho_{5,5}-\rho_{7,7})-\gamma\rho_{7,5},\
\ \dot{\rho}_{7,6}=-\gamma\rho_{7,6},\cr\cr
&&\dot{\rho}_{7,7}=i\Omega_{e_2}\rho_{7,1}-i\Omega_{e_1}^*\rho_{1,7}+i\Omega_{e_3}\rho_{7,5}-i\Omega_{e_3}^*\rho_{5,7}-\gamma\rho_{7,7},\
\ \dot{\rho}_{7,8}=-\gamma\rho_{7,8},\cr\cr
&&\dot{\rho}_{8,1}=-\gamma\rho_{8,1},\ \
\dot{\rho}_{8,2}=-\gamma\rho_{8,2},\ \
\dot{\rho}_{8,3}=-i\Omega_{e_2}^*(\rho_{3,3}-\rho_{8,8})-\gamma\rho_{8,3},\cr\cr
&&\dot{\rho}_{8,4}=-\gamma\rho_{8,4},\ \
\dot{\rho}_{8,5}=-\gamma\rho_{8,5},\ \
\dot{\rho}_{8,6}=-\gamma\rho_{8,6},\cr\cr
&&\dot{\rho}_{8,7}=-\gamma\rho_{8,7},\ \
\dot{\rho}_{8,8}=i\Omega_{e_2}\rho_{8,3}-i\Omega_{e_2}^*\rho_{3,8}-\gamma\rho_{8,8},\cr\cr
&&\dot{\rho}_{9,9}=\gamma\rho_{1,1}+\frac{\gamma}{2}(\rho_{5,5}+\rho_{7,7}),\
\
\dot{\rho}_{10,10}=\gamma\rho_{2,2}+\frac{\gamma}{2}(\rho_{6,6}+\rho_{7,7}),\cr\cr
&&\dot{\rho}_{11,11}=\gamma\rho_{3,3}+\frac{\gamma}{2}(\rho_{5,5}+\rho_{8,8}),\
\
\dot{\rho}_{12,12}=\gamma\rho_{4,4}+\frac{\gamma}{2}(\rho_{6,6}+\rho_{8,8}).
\end{eqnarray}
Assuming $\rho_{\ell,\ell'}=\tilde{\rho}_{\ell,\ell'}\exp(-\gamma
t)$, $(\ell,\ell'=1,2,...,8)$, we can find that the operator
$\tilde{\rho}(t)=\sum_{\ell,\ell'=1}^{8}\tilde{\rho}_{\ell,\ell'}|\Phi_\ell\rangle\langle\Phi_\ell'|$
satisfy the von Neumann equation
\begin{eqnarray}\label{apa2}
\dot{\tilde{\rho}}(t)=-i[H_e(t),\tilde{\rho}(t)].
\end{eqnarray}
Consequently, $\tilde{\rho}(t)$ can be calculated by
\begin{eqnarray}\label{apa3}
\tilde{\rho}(t)=U_e(t)\rho(0)U_e^\dag(t),
\end{eqnarray}
with $U_e(t)$ being the evolution operator given by the equation
$i\dot{U}_e(t)=H_e(t)U_e(t)$. Besides, we define
$\rho'(t)=\sum_{\ell=9}^{12}\tilde{\rho}_{\ell}|\Phi_\ell\rangle\langle\Phi_\ell|$.
The matrix elements of $\rho'(t)$ are given by
\begin{eqnarray}\label{apa4}
&&\rho_{9,9}(t)=\frac{\gamma}{2}\int_0^t[2\rho_{1,1}(t')+\rho_{5,5}(t')+\rho_{7,7}(t')]dt',\cr\cr
&&\rho_{10,10}(t)=\frac{\gamma}{2}\int_0^t[2\rho_{2,2}(t')+\rho_{6,6}(t')+\rho_{7,7}(t')]dt',\cr\cr
&&\rho_{11,11}(t)=\frac{\gamma}{2}\int_0^t[2\rho_{3,3}(t')+\rho_{5,5}(t')+\rho_{8,8}(t')]dt',\cr\cr
&&\rho_{12,12}(t)=\frac{\gamma}{2}\int_0^t[2\rho_{4,4}(t')+\rho_{6,6}(t')+\rho_{8,8}(t')]dt'.
\end{eqnarray}
Combining the results of Eqs.~(\ref{apa3}-\ref{apa4}), the density
operator $\rho(t)$ can be obtained as Eq.~(\ref{de2}).

\section*{Appendix C: Dynamic phase and geometric phase acquired in the implementation of single qubit gate}

We now prove that the dynamic (geometric) phase acquired in the time
interval $[0,\tau_1]$ is nullified by that acquired in time interval
$[\tau_2,T]$. Firstly, we calculate the time derivative of
$\beta_2(t)$ in time interval $[\tau_2,T]$ as
\begin{eqnarray}\label{apb1}
\frac{d}{dt}\beta_2(t)=\frac{d\tilde{t}}{dt}\frac{d}{d\tilde{t}}[-\Theta_s+\beta_2(\tilde{t})]=-\frac{\tau_1}{T-\tau_2}\times\frac{d}{d\tilde{t}}\beta_2(\tilde{t}),
\end{eqnarray}
with $\tilde{t}=\tau_1(T-t)/(T-\tau_2)$. According to
Eq.~(\ref{e16}), we have
\begin{eqnarray}\label{apb2}
&&\vartheta_-(T)-\vartheta_-(\tau_2)
=\int_{\tau_2}^{T}\frac{\dot{\beta}_2(t)
\sin^2[\beta_1(t)]}{2\cos[\beta_1(t)]}dt
=\int_{\tau_1}^{0}\frac{\dot{\beta}_2(\tilde{t})
\sin^2[\beta_1(\tilde{t})]}{2\cos[\beta_1(\tilde{t})]}d\tilde{t}
=-\vartheta_-(\tau_1),\cr\cr &&\Theta_-(T)-\Theta_-(\tau_2)
=-\int_{\tau_2}^{T}\dot{\beta}_2(t)\sin^2[\frac{\beta_1(t)}{2}]dt
=-\int_{\tau_1}^{0}\dot{\beta}_2(\tilde{t})\sin^2[\frac{\beta_1(\tilde{t})}{2}]d\tilde{t}
=-\Theta_-(\tau_1).\
\end{eqnarray}

\section*{Appendix D: Average fidelity}

We now make a brief introduction about the approach to calculate the
average fidelity proposed in Ref.~\cite{PedersenPLA367}. The theorem
in Ref.~\cite{PedersenPLA367} shows that, for any linear operator
$M$ on an $n$-dimensional complex Hilbert space, the uniform average
of $|\langle\psi|M|\psi\rangle|^2$ over state vectors $|\psi\rangle$
on the unit sphere $S^{2n-1}$ in $\mathbb{C}^n$ can be calculated by
\begin{eqnarray}\label{fpd1}
\int\limits_{S^{2n-1}}|\langle\psi|M|\psi\rangle|^2dV=\frac{1}{n(n+1)}[\mathrm{Tr}(MM^\dag)+|\mathrm{Tr}(M)|^2],
\end{eqnarray}
with $dV$ being the normalized measure on the sphere. Firstly, if
$M$ is a Hermitian operator, it can be diagonalized as a diagonal
operator $\Lambda$ by $\Lambda=\mathcal{U}M\mathcal{U}^-1$ via a
unitary operator $\mathcal{U}$. We denote the left-hand side and
right-hand side of Eq.~(\ref{fpd1}) as $\mathcal{L}(M)$ and
$\mathcal{R}(M)$, respectively. By a change of variables
$\psi\rightarrow\mathcal{U}\psi$, we obtain
$\mathcal{L}(\Lambda)=\mathcal{L}(\mathcal{U}M\mathcal{U}^-1)=\mathcal{L}(M)$.
On the other hand, considering the fact that
$\mathrm{Tr}(A_1A_2)=\mathrm{Tr}(A_2A_1)$ for two arbitrary linear
operators $A_1$ and $A_2$, we obtain
$\mathcal{R}(\Lambda)=\mathcal{R}(M)$. Since $\mathcal{L}(\Lambda)$
is a homogeneous polynomial of degree 2 in the real variables
$\lambda_1,\lambda_2,...,\lambda_n$, and unitary invariance implies
that it is invariant under the exchange of any two $\lambda_\jmath$
and $\lambda_{\jmath'}$ ($\jmath,\jmath'=1,2,...,n$), consequently
the only possible form of $\mathcal{L}(\Lambda)$ is
\begin{eqnarray}\label{fpd2}
\mathcal{L}(M)=\mathcal{L}(\Lambda)=a_1\mathrm{Tr}(\Lambda^2)+a_2\mathrm{Tr}^2(\Lambda)=a_1\mathrm{Tr}(MM^\dag)+a_2|\mathrm{Tr}(M)|^2,
\end{eqnarray}
with $a_1$ and $a_2$ being constants related to $n$. By considering
$M_x=|1\rangle\langle2|+|2\rangle\langle1|$,
$M_y=-i|1\rangle\langle2|+i|2\rangle\langle1|$,
$M_z=|1\rangle\langle1|-|2\rangle\langle2|$ and
$M_0=|1\rangle\langle1|+|2\rangle\langle2|$, we respectively derive
\begin{eqnarray}\label{fpd3}
&&\mathcal{L}(M_x)=2a_1=4\int\limits_{S^{2n-1}}\mathrm{Re}(c_1c_2^*)^2dV,\
\mathcal{L}(M_y)=2a_1=4\int\limits_{S^{2n-1}}\mathrm{Im}(c_1c_2^*)^2dV,\cr\cr&&
\mathcal{L}(M_z)=2a_1=\int\limits_{S^{2n-1}}(|c_1|^4+|c_2|^4-2|c_1|^2|c_2|^2)dV,\cr\cr&&
\mathcal{L}(M_0)=2a_1+4a_2=\int\limits_{S^{2n-1}}(|c_1|^4+|c_2|^4+2|c_1|^2|c_2|^2)dV,
\end{eqnarray}
with $|\psi\rangle=\sum_{\jmath=1}^n c_\jmath|\jmath\rangle$. From
Eq.~(\ref{fpd3}), we obtain
$\mathcal{L}(M_x)+\mathcal{L}(M_y)+\mathcal{L}(M_z)=\mathcal{L}(M_0)$,
which gives $a_1=a_2$. Furthermore, by picking the identity operator
$\mathbbold{1}$ in $\mathcal{L}(\Lambda)$, one can derive
$a_1n+a_2n^2=\mathcal{L}(\mathbbold{1})=1$. Combining the results
above, we can derive $a_1=a_2=1/n(n+1)$. Therefore, for a Hermitian
operator $M$, we have
$\mathcal{L}(M)=\mathcal{L}(\Lambda)=\mathcal{R}(\Lambda)=\mathcal{R}(M)$.

In fact, the result can also apply to an anti-Hermitian operator $A$
($A^\dag=-A$) since $L(A)=L(iA)=R(iA)=R(A)$. In addition, for a
general operator $M$, it can be decomposed into a Hermitian operator
$S=(M+M^\dag)/2$ and an anti-Hermitian operator $A=(M-M^\dag)/2$
with $M=S+A$. Accordingly, one can derive
\begin{eqnarray}\label{fpd4}
\mathcal{L}(S+A)&=&\mathcal{L}(S)+\mathcal{L}(A)+\int\limits_{S^{2n-1}}\langle\psi|S|\psi\rangle\langle\psi|(A+A^\dag)|\psi\rangle=\mathcal{L}(S)+\mathcal{L}(A),\cr\cr
\mathcal{R}(S+A)&=&\frac{1}{n(n+1)}\{\mathrm{Tr}[(S+A)(S-A)]+[\mathrm{Tr}(S)+\mathrm{Tr}(A)][\mathrm{Tr}(S)-\mathrm{Tr}(A)]\}\cr\cr
&=&\mathcal{R}(S)+\mathcal{R}(A).
\end{eqnarray}
Using the results of $\mathcal{L}(S)=\mathcal{R}(S)$ and
$\mathcal{L}(A)=\mathcal{R}(A)$, we have the result of
Eq.~(\ref{fpd1}) is satisfied for arbitrary linear operator $M$.
Specially, for quantum gates in a considered
$\mathcal{N}$-dimensional computational subspace $\mathcal{S}_c$
(with projection operator $\mathcal{P}_c$), assuming the target
operation and real evolution are described by unitary operators
$\tilde{U}$ and $U(t)$, the average fidelity over all possible
initial state $|\psi\rangle$ in the subspace $\mathcal{S}_c$ should
be
\begin{eqnarray}\label{fpd5}
\bar{F}(t)=\int\limits_{S^{2\mathcal{N}-1}}|\langle\psi|\mathcal{P}_c\tilde{U}^\dag
U(t)\mathcal{P}_c|\psi\rangle|^2dV,
\end{eqnarray}
with $S^{2\mathcal{N}-1}$ being the unit sphere of the computational
subspace $\mathcal{S}_c$. By substituting
$M=\mathcal{P}_c\tilde{U}^\dag U(t)\mathcal{P}_c$, we obtain the
formula to calculate the average fidelity used in Eqs.~(\ref{e22})
and (\ref{e28}).

\section*{Appendix E: Possible extensions of the protocol}

\begin{figure}
\scalebox{0.6}{\includegraphics[scale=1]{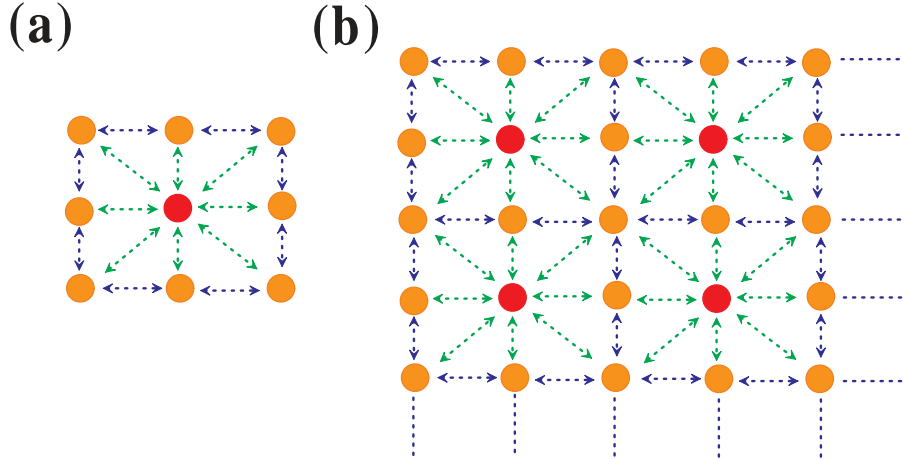}} \caption{(a)
Diagrammatic sketch of a possible structure for realizing
large-scale quantum computation with single auxiliary atom. (b)
Diagrammatic sketch of a possible structure for realizing
large-scale quantum computation with multiple auxiliary atom in an
atom array.}\label{fig6}
\end{figure}

We now discuss the possible extensions of the protocol to
large-scale quantum computation. First, we consider a structure
shown in Fig.~\ref{fig6}(a), where the red dot in the middle
represents an auxiliary atom, and the orange dots around the red dot
are the computational atoms. Theoretically, we can realize
single-qubit gates for any computational atom with a single
auxiliary atom if every atom in this system have enough strong
Rydberg interaction strength with the auxiliary atom. In addition,
if two adjacent computational atoms have enough strong Rydberg
interaction, we can realize two-qubit gates of them with a single
auxiliary atom. In the ideal case, the computational atom remain in
their ground states at the beginning and the end of a gate
implementation. Therefore, the computational atoms without laser
driving would not influence the computational atom being
manipulated. Moreover, if the heralded implementation is successful,
the auxiliary atom will return to its Rydberg state $|r\rangle$. In
this case, the auxiliary atom can be continued to use in the next
step of operations. If the measurement result shows that the
auxiliary atom is in its ground state $|0\rangle$, the
implementation of quantum gate is failed, and we need to initialize
the auxiliary atom to the Rydberg state $|r\rangle$ again. With the
structure in Fig.~\ref{fig6}(a), it is possible to realize
single-qubit gates of all computational atoms and realize two-qubit
gates of arbitrary pairs of adjacent atoms. In principle, we may add
many computational atoms around the auxiliary atom, but trapping,
distant control and addressing of atoms may become difficult if the
number of computational atoms are very large. Therefore, to realize
large-scale quantum computation, the atom array shown in
Fig.~\ref{fig6}(b) may be an alternative candidate. As shown in
Fig.~\ref{fig6}(b), the atom array is composed of many repeated
blocks of atom shown in Fig.~\ref{fig6}(a). The atom in the middle
of each block can be used as an auxiliary atom to implement heralded
quantum gates. To date, the manipulation of Rydberg atom array have
been studied in several previous works
\cite{SamajdarPRL124,LJCPRB101}, and many interesting results are
shown. Therefore, the Rydberg atom array may be a promising platform
for large-scale quantum computation.


\begin{thebibliography}{999}
\bibitem{note1} By applying pulses with contrary-sign detunings, we can offset the Stark-shifts of ground
states. For example, considering $\tilde{H}_{1}(t)$ in
Eq.~(\ref{fp3}), the Stark shifts caused by the pulses with blue
detunings are described by
$H_{ss1}(t)=\frac{|\Omega_1(t)|^2}{\Delta_1}(|0++\rangle_{012}\langle0++|+|0+-\rangle_{012}\langle0+-|-|0r+\rangle_{012}\langle0r+|-|0r-\rangle_{012}\langle0r-|)
+\frac{|\Omega_2(t)|^2}{\Delta_1}(|0++\rangle_{012}\langle0++|+|0-+\rangle_{012}\langle0-+|-|0+r\rangle_{012}\langle0+r|-|0-r\rangle_{012}\langle0-r|)$.
While for Stark shifts caused by the pulses with red detunings, they
can be described by $-H_{ss1}(t)$. Thus, the total Stark shifts on
ground states $|0\pm\pm\rangle_{0,1,2}$ are offset. By calculating
the Stark shifts with $\tilde{H}_{2}(t)$ and $\tilde{H}_{3}(t)$ in
Eq.~(\ref{fp3}), we can obtain similar results.
\bibitem{note2} Here, the Rydberg interaction strengths are considered the
same for the simplicity of discussions. In fact, the Rydberg
interaction strengths between each pair of atoms can be different.
We can see from Eq.~(\ref{fp3}) that terms with oscillation
frequencies in the scale of $V$ are omitted. Therefore, when the
Rydberg interaction strengths become different, as long as they are
much larger compared with detunings and Rabi frequencies of laser
pulses, we can obtain the same result as Eq.~(\ref{fp3}).
\end{thebibliography}
\end{document}